\documentclass[aps,prd,reprint,showpacs, preprintnumbers,amsmath,amssymb,superscriptaddress, nofootinbib]{revtex4-1}
\usepackage{amssymb, amsmath, braket, nicefrac, graphicx} 
\usepackage{relsize}
\usepackage{lineno}
\usepackage{setspace}

\usepackage{esint}
\usepackage{chngpage}
\usepackage{epsfig}
\usepackage{multirow}
\usepackage{rotating}
\usepackage{float}
\usepackage{color}
\usepackage{xcolor}
\usepackage{appendix}
\usepackage{ amstext }
\usepackage{ gensymb }
\usepackage{ textcomp }
\usepackage{ wasysym }
\usepackage{pstricks}

\usepackage[caption=false,labelformat=parens]{subfig}

\usepackage{dcolumn}
\usepackage{bm}
\usepackage{units}

\newcommand{\enu}{\mbox{$E_\nu$}}

\newcommand{\numu}{\mbox{$\nu_{\mu}$}}                         
\newcommand{\nue}{\mbox{$\nu_{e}$}}                                

\newcommand{\anu}{\ensuremath{\bar{\nu}}}

\newcommand{\anue}{\mbox{$\overline{\nu}_{e}$}}                      

\newcommand{\anumu}{\ensuremath{\bar{\nu}_{\mu}}}

\newcommand{\simgt}{\,\hbox{\lower0.6ex\hbox{$\sim$}\llap{\raise0.6ex\hbox{$>$}}}\,}
\newcommand{\simlt}{\,\hbox{\lower0.6ex\hbox{$\sim$}\llap{\raise0.6ex\hbox{$<$}}}\,}

\newcommand{\numi}{NuMI}

\newcommand{\eavail}{$E_{\rm{avail}}$}
\newcommand{\qthree}{$|\vec{q}\,|$}

\definecolor{maroon}{RGB}{162,10,10}

\usepackage{hyperref}
\usepackage[all]{hypcap}
\hypersetup{
    colorlinks,
    citecolor=black,
    filecolor=black,
    linkcolor=black,
    urlcolor=black
}

\makeatletter
\renewenvironment{table}
  {\def\@captype{table}}
  {}

\renewenvironment{figure}
  {\def\@captype{figure}}
  {}
\makeatother
   
\begin{document}
\preprint{FERMILAB-PUB-24-0571-PPD}
\title{Measurement of $d^{2}\sigma/d |\vec{q}| dE_{\rm{avail}}$ in charged current $\numu$-nucleus interactions 
at $\left<E_{\nu}\right>$ = 1.86 GeV using the NOvA Near Detector} 

\newcommand{\ANL}{Argonne National Laboratory, Argonne, Illinois 60439, 
USA}
\newcommand{\Bandirma}{Bandirma Onyedi Eyl\"ul University, Faculty of 
Engineering and Natural Sciences, Engineering Sciences Department, 
10200, Bandirma, Balıkesir, Turkey}
\newcommand{\ICS}{Institute of Computer Science, The Czech 
Academy of Sciences, 
182 07 Prague, Czech Republic}
\newcommand{\IOP}{Institute of Physics, The Czech 
Academy of Sciences, 
182 21 Prague, Czech Republic}
\newcommand{\Atlantico}{Universidad del Atlantico,
Carrera 30 No.\ 8-49, Puerto Colombia, Atlantico, Colombia}
\newcommand{\BHU}{Department of Physics, Institute of Science, Banaras 
Hindu University, Varanasi, 221 005, India}
\newcommand{\UCLA}{Physics and Astronomy Department, UCLA, Box 951547, Los 
Angeles, California 90095-1547, USA}
\newcommand{\Caltech}{California Institute of 
Technology, Pasadena, California 91125, USA}
\newcommand{\Cochin}{Department of Physics, Cochin University
of Science and Technology, Kochi 682 022, India}
\newcommand{\Charles}
{Charles University, Faculty of Mathematics and Physics,
 Institute of Particle and Nuclear Physics, Prague, Czech Republic}
\newcommand{\Cincinnati}{Department of Physics, University of Cincinnati, 
Cincinnati, Ohio 45221, USA}
\newcommand{\CSU}{Department of Physics, Colorado 
State University, Fort Collins, CO 80523-1875, USA}
\newcommand{\CTU}{Czech Technical University in Prague,
Brehova 7, 115 19 Prague 1, Czech Republic}
\newcommand{\Dallas}{Physics Department, University of Texas at Dallas,
800 W. Campbell Rd. Richardson, Texas 75083-0688, USA}
\newcommand{\DallasU}{University of Dallas, 1845 E 
Northgate Drive, Irving, Texas 75062 USA}
\newcommand{\Delhi}{Department of Physics and Astrophysics, University of 
Delhi, Delhi 110007, India}
\newcommand{\JINR}{Joint Institute for Nuclear Research,  
Dubna, Moscow region 141980, Russia}
\newcommand{\Erciyes}{
Department of Physics, Erciyes University, Kayseri 38030, Turkey}
\newcommand{\FNAL}{Fermi National Accelerator Laboratory, Batavia, 
Illinois 60510, USA}
\newcommand{\FSU}{Florida State University, Tallahassee, Florida 32306, USA}
\newcommand{\UFG}{Instituto de F\'{i}sica, Universidade Federal de 
Goi\'{a}s, Goi\^{a}nia, Goi\'{a}s, 74690-900, Brazil}
\newcommand{\Guwahati}{Department of Physics, IIT Guwahati, Guwahati, 781 
039, India}
\newcommand{\Harvard}{Department of Physics, Harvard University, 
Cambridge, Massachusetts 02138, USA}
\newcommand{\Houston}{Department of Physics, 
University of Houston, Houston, Texas 77204, USA}
\newcommand{\IHyderabad}{Department of Physics, IIT Hyderabad, Hyderabad, 
502 205, India}
\newcommand{\Hyderabad}{School of Physics, University of Hyderabad, 
Hyderabad, 500 046, India}
\newcommand{\IIT}{Illinois Institute of Technology,
Chicago IL 60616, USA}
\newcommand{\Indiana}{Indiana University, Bloomington, Indiana 47405, 
USA}
\newcommand{\INR}{Institute for Nuclear Research of Russia, Academy of 
Sciences 7a, 60th October Anniversary prospect, Moscow 117312, Russia}
\newcommand{\Iowa}{Department of Physics and Astronomy, Iowa State 
University, Ames, Iowa 50011, USA}
\newcommand{\Irvine}{Department of Physics and Astronomy, 
University of California at Irvine, Irvine, California 92697, USA}
\newcommand{\Jammu}{Department of Physics and Electronics, University of 
Jammu, Jammu Tawi, 180 006, Jammu and Kashmir, India}
\newcommand{\Lebedev}{Nuclear Physics and Astrophysics Division, Lebedev 
Physical 
Institute, Leninsky Prospect 53, 119991 Moscow, Russia}
\newcommand{\Magdalena}{Universidad del Magdalena, Carrera 32 No 22-08 Santa Marta, Colombia}
\newcommand{\MSU}{Department of Physics and Astronomy, Michigan State 
University, East Lansing, Michigan 48824, USA}
\newcommand{\Crookston}{Math, Science and Technology Department, University 
of Minnesota Crookston, Crookston, Minnesota 56716, USA}
\newcommand{\Duluth}{Department of Physics and Astronomy, 
University of Minnesota Duluth, Duluth, Minnesota 55812, USA}
\newcommand{\Minnesota}{School of Physics and Astronomy, University of 
Minnesota Twin Cities, Minneapolis, Minnesota 55455, USA}
\newcommand{\Mississippi}{University of Mississippi, University, Mississippi 38677, USA}
\newcommand{\NISER}{National Institute of Science Education and Research,
Khurda, 752050, Odisha, India}
\newcommand{\Oxford}{Subdepartment of Particle Physics, 
University of Oxford, Oxford OX1 3RH, United Kingdom}
\newcommand{\Panjab}{Department of Physics, Panjab University, 
Chandigarh, 160 014, India}
\newcommand{\Pitt}{Department of Physics, 
University of Pittsburgh, Pittsburgh, Pennsylvania 15260, USA}
\newcommand{\QMU}{Particle Physics Research Centre, 
Department of Physics and Astronomy,
Queen Mary University of London,
London E1 4NS, United Kingdom}
\newcommand{\RAL}{Rutherford Appleton Laboratory, Science 
and 
Technology Facilities Council, Didcot, OX11 0QX, United Kingdom}
\newcommand{\SAlabama}{Department of Physics, University of 
South Alabama, Mobile, Alabama 36688, USA} 
\newcommand{\Carolina}{Department of Physics and Astronomy, University of 
South Carolina, Columbia, South Carolina 29208, USA}
\newcommand{\SDakota}{South Dakota School of Mines and Technology, Rapid 
City, South Dakota 57701, USA}
\newcommand{\SMU}{Department of Physics, Southern Methodist University, 
Dallas, Texas 75275, USA}
\newcommand{\Stanford}{Department of Physics, Stanford University, 
Stanford, California 94305, USA}
\newcommand{\Sussex}{Department of Physics and Astronomy, University of 
Sussex, Falmer, Brighton BN1 9QH, United Kingdom}
\newcommand{\Syracuse}{Department of Physics, Syracuse University,
Syracuse NY 13210, USA}
\newcommand{\Tennessee}{Department of Physics and Astronomy, 
University of Tennessee, Knoxville, Tennessee 37996, USA}
\newcommand{\Texas}{Department of Physics, University of Texas at Austin, 
Austin, Texas 78712, USA}
\newcommand{\Tufts}{Department of Physics and Astronomy, Tufts University, Medford, 
Massachusetts 02155, USA}
\newcommand{\UCL}{Physics and Astronomy Department, University College 
London, 
Gower Street, London WC1E 6BT, United Kingdom}
\newcommand{\Virginia}{Department of Physics, University of Virginia, 
Charlottesville, Virginia 22904, USA}
\newcommand{\WSU}{Department of Mathematics, Statistics, and Physics,
 Wichita State University, 
Wichita, Kansas 67260, USA}
\newcommand{\WandM}{Department of Physics, William \& Mary, 
Williamsburg, Virginia 23187, USA}
\newcommand{\Wisconsin}{Department of Physics, University of 
Wisconsin-Madison, Madison, Wisconsin 53706, USA}
\newcommand{\deceased}{Deceased.}
\affiliation{\ANL}
\affiliation{\Atlantico}
\affiliation{\Bandirma}
\affiliation{\BHU}
\affiliation{\Caltech}
\affiliation{\Charles}
\affiliation{\Cincinnati}
\affiliation{\Cochin}
\affiliation{\CSU}
\affiliation{\CTU}
\affiliation{\Delhi}
\affiliation{\Erciyes}
\affiliation{\FNAL}
\affiliation{\FSU}
\affiliation{\UFG}
\affiliation{\Guwahati}
\affiliation{\Houston}
\affiliation{\Hyderabad}
\affiliation{\IHyderabad}
\affiliation{\IIT}
\affiliation{\Indiana}
\affiliation{\ICS}
\affiliation{\INR}
\affiliation{\IOP}
\affiliation{\Iowa}
\affiliation{\Irvine}
\affiliation{\JINR}
\affiliation{\Magdalena}
\affiliation{\MSU}
\affiliation{\Duluth}
\affiliation{\Minnesota}
\affiliation{\Mississippi}
\affiliation{\NISER}
\affiliation{\Panjab}
\affiliation{\Pitt}
\affiliation{\QMU}
\affiliation{\SAlabama}
\affiliation{\Carolina}
\affiliation{\SMU}
\affiliation{\Sussex}
\affiliation{\Syracuse}
\affiliation{\Texas}
\affiliation{\Tufts}
\affiliation{\UCL}
\affiliation{\Virginia}
\affiliation{\WSU}
\affiliation{\WandM}
\affiliation{\Wisconsin}

\author{M.~A.~Acero}
\affiliation{\Atlantico}

\author{B.~Acharya}
\affiliation{\Mississippi}

\author{P.~Adamson}
\affiliation{\FNAL}



\author{L.~Aliaga}
\affiliation{\FNAL}






\author{N.~Anfimov}
\affiliation{\JINR}


\author{A.~Antoshkin}
\affiliation{\JINR}


\author{E.~Arrieta-Diaz}
\affiliation{\Magdalena}

\author{L.~Asquith}
\affiliation{\Sussex}


\author{A.~Aurisano}
\affiliation{\Cincinnati}


\author{A.~Back}
\affiliation{\Indiana}
\affiliation{\Iowa}



% ...

\collaboration{The NOvA Collaboration}
\noaffiliation


\date{\today}

\begin{abstract}
Double- and single-differential cross sections for 
inclusive charged-current \numu-nucleus scattering are reported for the
kinematic domain 0 to 2 GeV/$c$ in three-momentum transfer and 0 to 2 GeV in available energy, 
at a mean $\numu$ energy of 1.86 GeV.   The measurements are based on an estimated 
995,760 \numu\, charged-current (CC) interactions in the scintillator medium of the NOvA Near Detector.
The subdomain populated by 2-particle-2-hole reactions is identified by the
cross-section excess relative to predictions for \numu-nucleus scattering that are constrained by a data
control sample.   Models for 2-particle-2-hole processes are rated by $\chi^2$ comparisons of the 
predicted-versus-measured \numu\,CC inclusive cross section over the full phase space
and in the restricted subdomain.  Shortfalls are observed in neutrino generator predictions 
obtained using the theory-based Val\`{e}ncia and SuSAv2 2p2h models.
\end{abstract}

\maketitle


\section{Introduction}
\label{sec:Intro}
 A dedicated campaign is underway by the neutrino physics community to 
 obtain a comprehensive picture of charged-current neutrino-nucleus
 interactions in the sub-GeV to few-GeV region of incident neutrino energies.   
 Through the first decade of the present millenium, treatments of 
exclusive-channel neutrino scattering were largely based on 
hydrogen and deuterium bubble chamber data~\cite{Rein--Sehgal-1981, Gallagher-ARNP-2011}.   
The high-statistics neutrino-nucleus experiments of more recent times 
have resulted in refinements to the modeling of CC quasielastic scattering (CCQE) and of
baryon-resonance production (RES) initiated 
by $\nu/\anu$-nucleus scattering~\cite{Morfin:2012kn, Katori-Martini-2018}.   
Shallow and deep inelastic CC scattering (DIS) have also received renewed scrutiny
and modeling refinements~\cite{Morfin-2021}.  
Similarly, various aspects of neutrino CC coherent scattering (COH)
and of kaon and hyperon production have been clarified~\cite{Alvarez-Ruso-2014}.  
The emerging theme from these developments is that
neutrino-nucleus scattering involves much more than just neutrino-nucleon scattering 
in a relativistic Fermi gas.  The presence of a nuclear medium introduces new phenomena
whose observational effects must be understood to complete the picture of neutrino-nucleus interactions.

Study of neutrino-nucleus scattering receives strong impetus from neutrino
oscillation experiments as continued progress requires precise knowledge of
differential cross sections.  Neutrino oscillation measurements
provide a window into the underlying physics and 
symmetries of neutrino states.   At present, the ordering of neutrino mass eigenstates is unknown,
the extent to which charge conjugation plus parity (CP) symmetry 
is violated in the lepton sector remains to be ascertained, and the 
octant assignment for the flavor mixing angle $\theta_{23}$ -- if indeed it deviates from maximal 
mixing ($45^{\circ}$) -- needs to be resolved~\cite{MINOS+2020, T2K-Oscillations-2021, NOvA-Oscillations-2021}.
More precise knowledge of neutrino and antineutrino interactions in nuclear environments is 
required for experimental clarification of these fundamental questions.

A notable recent realization is that neutrino event rates in the sub- to few-GeV range of neutrino energy,
$E_{\nu}$, used by many of the oscillation experiments, receive contributions from multinucleon
initial states.  That interactions may involve two initial-state nucleons 
was known from electron-nucleus scattering~\cite{Ericson-1984}.
However, the possibility that similar excitations occur in neutrino scattering, though
mentioned in a 1985 paper by Delorme and Ericson~\cite{Ericson-1985}, was 
not generally recognized for some time.  Initial hints in neutrino data
came in the guise of unusually high values inferred for the axial mass parameter, $M_{A}$, 
of the axial-vector form factor, obtained with high-statistics samples of 
$\numu$-nucleus CC quasielastic-like scattering.  
In 2006--7, the K2K experiment reported $M_{A}$ to be 1.20$\pm$0.12 GeV from neutrinos 
on oxygen~\cite{K2KMA:2006} and subsequently 1.14$\pm$0.11 GeV 
for neutrinos on carbon~\cite{Scibar:2007}.  At the time, the world-average axial-vector mass 
hovered around 1.00 GeV/$c^2$ with uncertainty of $\sim$1\%~\cite{Rochester-2008, Kuzmin-2008}.   
Thus it came as a shock during 2008--10 when MiniBooNE, 
presenting new studies of neutrino CCQE interactions
in a carbon medium~\cite{MiniBooNEMA:2008, MiniBooNEMA:2010},
reported the ``effective value" of quasielastic $M_{A}$ to be 
$1.35 \pm 0.17$ GeV~\cite{MiniBooNEMA:2010}.  
High values for $M_{A}$ reflect the presence of an additional
reaction rate above that expected from neutrino scattering on quasi-free nucleons.   
That the data exhibit this feature has been abundantly confirmed 
in measurements by MiniBooNE~\cite{MiniBooNEMA:2013}, 
MINOS~\cite{MINOS-2015}, MINERvA~\cite{Rodrigues-2016, Gran-2018, Ascencio-2022}, 
T2K~\cite{T2K-2018, T2K-2023}, MicroBooNE~\cite{M-BooNE-2020, M-BooNE-2023}, and NOvA~\cite{Adjust-Models-2020}.  
The apparent high values for effective $M_{A}$ 
in $\numu$-nucleus CCQE interactions were driven by the omission 
in the analyses of so-called 2-particle 2-hole (2p2h) processes:
\begin{equation}
\label{reaction-1}
\nu_{\mu} + \mathcal{A}_{(nN + \mathcal{A}')}  \rightarrow \mu^{-} + p + N + \mathcal{A}',
\end{equation}
\noindent
where $n$, $p$, and $N$ designate a neutron, proton, and nucleon (either a neutron or proton), respectively.
Here, the incident neutrino interacts with nucleus $\mathcal{A}$ to
give a muon, proton, and nucleon in the final state.   The remnant nucleus $\mathcal{A}'$ 
with two holes in its Fermi sea subsequently undergoes deexcitation with possible nucleon ejection.

Theoretical calculations by the Lyon group were the first to explain the anomalous 
MiniBooNE CCQE result as originating with N-particle-N-hole interactions involving more than one nucleon, 
with N=2 giving the dominant contribution~\cite{Martini:2009uj, Martini:2010ex, Martini:2011wp}.   
Soon thereafter the Val\`{e}ncia group presented a detailed N-particle-N-hole model with 2p2h giving the dominant 
contribution~\cite{Nieves:2011pp, Nieves:2012pl, Gran-PRD-2013, Valencia-2020}.   
Both of these microscopic models utilize the graphs and calculational methods of many-body quantum field theory. 
More recently, models of somewhat different construction have been presented.  For example,
the SuSAv2 model uses superscaling (SuSA), an approximation that invokes universal scaling functions 
for the electromagnetic and weak interactions, to describe single-body nuclear effects.  
In SuSAv2 this superscaling, together with microscopic 
calculations based on meson-exchange current (MEC) diagrams, are incorporated into
a fully relativistic framework~\cite{Megias:2015, Simo-2016, SuSAv2-carbon-2022}.
Additionally, semi-empirical approaches have been implemented in the GENIE~\cite{Dytman-MEC} and
GiBUU~\cite{Gallmeister-2016, Mosel-2019} neutrino event generators.    
In paragraphs and figures to follow, the acronym ``2p2h" refers to the full suite of 
multinucleon processes invoked by the models.

In recent times, phenomenological predictions
have been probed at new levels of detail by 
detector-resolution-unfolded, double-differential (or even triple-differential) cross-section measurements.   
Initially this approach was applied to $\numu$ and $\anumu$ 
quasielastic-like scattering~\cite{ddXsec-MiniBooNE, ddXsec-T2K, ddXsec-AQE-MINERvA, Ruterbories-QE-like}.
More recently it has been used to characterize CC inclusive cross sections 
as well~\cite{ddXsec-Inc-MINERvA, NOvA-CC-inclusive}.   
The latter measurements are generally restricted to final-state muon kinematic variables, either
to muon production angle and kinetic energy, or to muon transverse and longitudinal momenta.  
Exceptions to this were two MINERvA investigations of nuclear-medium effects 
for $\numu$-carbon and $\anumu$-carbon scattering~\cite{Rodriques-2016, Gran-2018}  that reported
double-differential cross sections using three-momentum transfer, $|\vec{q}\,|$, and available energy, $E_{\rm{avail}}$.   
The $E_{\rm{avail}}$ variable represents final-state hadronic energy that is capable of
producing ionization in the detector;  it is the sum of electron, proton, charged pion, 
and kaon kinetic energy, plus neutral pion and photon total energy.   
For hyperons, $E_{\rm{avail}}$ is
the total energy minus the nucleon mass; for antinucleons it is the total energy including rest mass.
Available energy as used here excludes energies initiated by neutrons, as neutron scattering
mostly does not register in detectors that rely on scintillation in hydrocarbons.
Available energy is useful as a proxy for energy transfer, $q_{0}$, in CC interactions 
because it minimizes detector-specific, model-dependent corrections that reconstruction of $q_{0}$
requires for unobserved energies.

The main motivation for choosing $E_{\rm{avail}}$ and reconstructed $|\vec{q}\,|$ is that they are 
experimental observables that closely resemble ($q_{0}, |\vec{q}\,|)$, the latter being the natural
variables for the nuclear physics phenomenology associated with 2p2h~\cite{Gran-PRD-2013, Gallmeister-2016}. 
 Assuming that the prevailing picture of 2p2h is roughly correct,  
$\numu$ scattering on nucleon pairs results in energetic $pp$ or $pn$ pairs
appearing in the final state.   Then $|\vec{q}\,|$ of de Broglie wavelength 
$\leq$ 4\,fm (corresponding to $|\vec{q}\,| >$ 0.3\,GeV/$c$) is well-suited to probe the initial state,
while $E_{\rm{avail}}$ measures the energy transfer to the target system.

This work uses data recorded by the NOvA Near Detector to measure the double-differential
cross section in $|\vec{q}\,|$ and $E_{\rm{avail}}$ of \numu\,CC inclusive interactions
\begin{equation}
\label{reaction-inclusive}
\nu_{\mu} (k) + \mathcal{A}  \rightarrow \mu^{-}(k')  + X.
\end{equation}
Here, $k$ and $k'$ are the four-momenta of the incident neutrino and the outgoing muon.   
The NOvA data provide a high-statistics sample of neutrino CC interactions in a hydrocarbon-rich medium, 
concentrated in an $E_{\nu}$ range from 1.0\,GeV to 5.0\,GeV.  
This range overlaps but mostly lies above the $E_{\nu}$ range analyzed by T2K~\cite{T2K-2018, T2K-2023}.
The NOvA sample also overlaps, but provides more focus on, the low end of the $E_{\nu}$ range
$\sim1 < E_{\nu} < 20$ GeV examined by MINERvA~\cite{Rodrigues-2016, Ascencio-2022}.   
Compared to existing and future studies of neutrino interactions in argon, the NOvA data lies mostly above the
0.2\,GeV to 1.5\,GeV $E_{\nu}$ range targeted by MicroBooNE~\cite{MicroB-2024} 
and by the SBND and ICARUS experiments~\cite{SBN-Proposal} 
while covering the lower half of the high-flux plateau in the $\numu$ energy spectrum 
planned for the DUNE experiment~\cite{DUNE-2020}.

\section{Neutrino Beam, Near Detector, and Data Exposure}
\label{sec:ND-nf-de}
\begin{figure}
\begin{centering}
\includegraphics[scale=0.48]{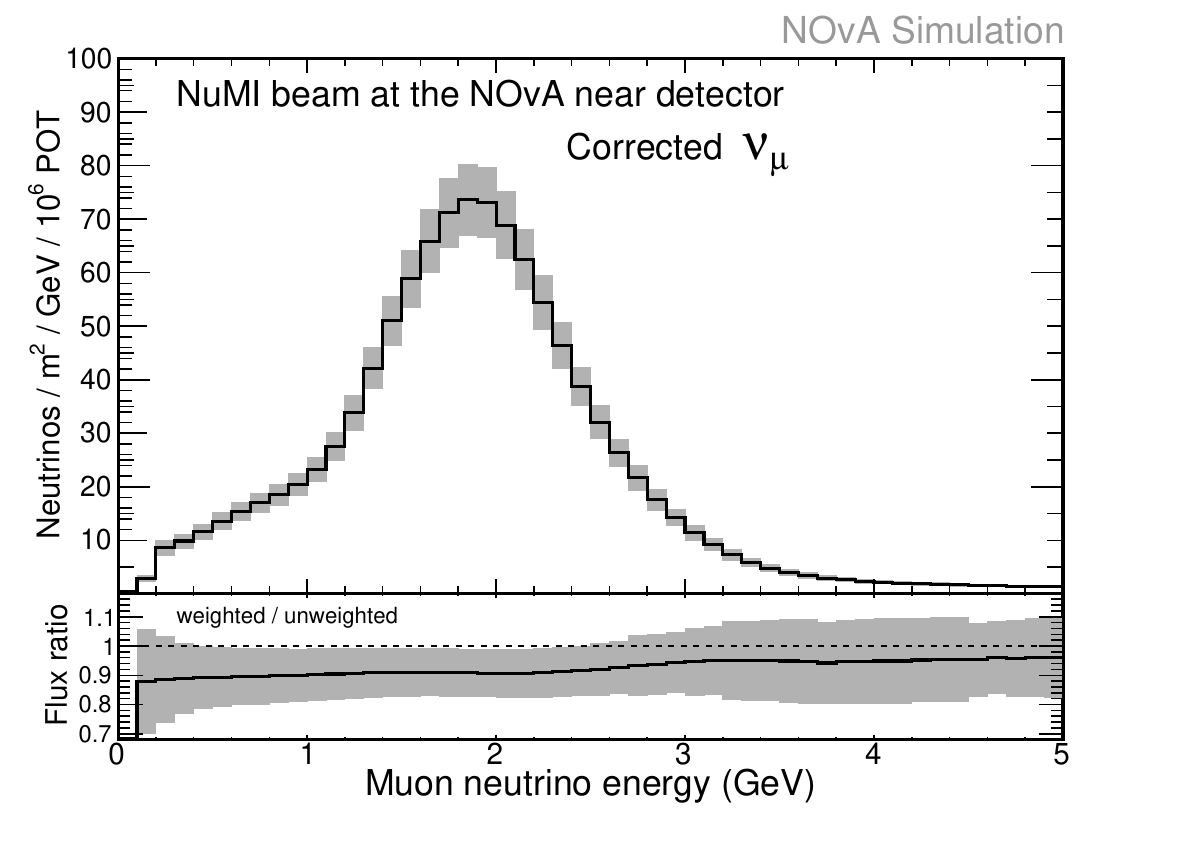}
\par\end{centering}
\caption{\label{Fig01} The flux of $\numu$ neutrinos at the NOvA Near Detector (solid histogram)
shown with the total 1-$\sigma$ uncertainty (shaded band)~\cite{NOvA-CC-inclusive}.}
 \end{figure}

The NuMI neutrino beam at Fermilab~\cite{NuMI-Beam-2016}
 is produced by directing 120\,GeV protons from the Main Injector accelerator 
onto a 1.2-m-long graphite target. Charged hadrons produced in the target traverse two
magnetic focusing horns that are positioned immediately downstream.
Operation of the horns in the forward horn-current mode results in focusing of 
positively charged pions and kaons.   
These positive mesons are then directed into a 675-m-long drift region where they decay 
to produce antimuons and muon neutrinos.  The resulting $\numu$ flux is calculated 
using a detailed simulation of beamline components and of the hadronic shower 
that emerges from the graphite target and evolves 
into mesons decaying to neutrinos.    The  simulation is based on Geant4 v9.2.p03 
with the FTFP BERT hadron production model~\cite{Geant4-2003, Geant4-2006, Geant4-2016}.  
The PPFX package~\cite{NuMI-Flux-Aliaga-2016}
is used to adjust the hadronic model to bring it into agreement
with constraints provided by external hadron production data~\cite{Paley-MIPP, Alt-NA49, Abgrall-NA61, Barton-83, Seun-07, 
Lebedev-07, Tinti-10, Baatar-NA49, Skubic-78, Denisov-73, Carroll-79, Abe-13-13, 
Cronin-57, Allaby-69, Longo-62, Bobchenko-79, Fedorov-78, Abrams-70}.   
In the neutrino energy range relevant to this measurement (1.0\,-\,5.0 GeV) 
and at the NOvA off-axis angle of 14.6\,mrad, 97.5\% of the NuMI
forward horn-current neutrino flux consists of $\numu$ neutrinos.
The remainder includes a 1.8\% $\anumu$ component arising from decay of negatively charged mesons.   
There is also a contribution from $\nue$ and $\anue$ neutrinos of 0.7\% 
in this energy range~\cite{NewConstraints-2018, NOvA-CC-inclusive}.    
The $\numu$ neutrino flux spectrum predicted at the ND is shown in Fig.~\ref{Fig01}.  The
ratio shown in the lower panel depicts the flux correction using PPFX~\cite{NuMI-Flux-Aliaga-2016} 
with respect to the FTFP BERT hadronic model of Geant4 v9.2.p03. 

The analyzed $\numu$-nucleus interactions occurred 
in the liquid scintillator tracking medium of the NOvA Near Detector (ND)~\cite{NOvA-TDR}. 
The ND is a 193-ton active mass, segmented tracking calorimeter 
located 100\,m underground.  It is constructed from 
polyvinyl chloride cells of rectangular-prism shape (length = 3.9\,m, width = 3.9\,cm, 6.6\,cm 
depth in beam direction) which are extruded together 
in units and joined along the long edges to form square planes of 96 cells per plane~\cite{Talaga-2017}.   
The cells are filled with organic liquid scintillator with trace concentrations of wavelength-shifting fluors~\cite{Mufson-2015}.   
The planes are aligned transverse to the beam direction in alternating horizontal and vertical orientations, 
enabling three-dimensional event reconstruction with $\sim$4\,cm granularity in the transverse dimensions.   
The active volume consists of 
192 contiguous planes extending 12.7\,m along the beam direction.  It presents a target medium 
made of 63\% active material with a radiation length of 38\,cm, 
whose nuclear composition consists of carbon (66.7\% by mass), chlorine (16.1\%), 
hydrogen (10.8\%), titanium (3.2\%), oxygen (3.0\%), and other nuclei ($<0.3$\%)~\cite{NOvA-CC-inclusive}.   

The downstream end of the ND is outfitted with a ``muon catcher.''  
It consists of 10-cm-thick steel planes stacked along the beam direction, 
each of which is sandwiched between a pair of scintillator planes.   Within the pair,
one plane is vertically oriented and the other is horizontally oriented.   
The entire sequence contains ten steel planes and eleven pairs of scintillator planes. 
Including the muon catcher and scintillator tracking volumes together, the ND is capable of stopping muons of 
kinetic energy up to 2.5 GeV.

Scintillation light produced by traversal of charged particles 
through a cell of the ND is collected via a loop of wavelength-shifting optical fiber 
and routed to an avalanche photodiode (APD) at the end of the cell.   
The APD signals are continuously digitized, and those that exceed a noise-vetoing threshold 
are sent to a data buffer.   Receipt of a time stamp from the Fermilab accelerator 
prior to the delivery of the 10 $\mu$s beam spill initiates 
the recording of a 550 $\mu$s portion of data (that includes the beam spill), which is saved for analysis.

A detailed model of the ND, together with a combination 
of Geant4 v10.1.p03 with the QGSP BERT HP 
hadron production model~\cite{Geant4-2003, Geant4-2006, Geant4-2016} 
and custom software,
is used to simulate the detector's response 
to particles initiated by individual interactions.   The simulation, which is
tuned to reproduce measured scintillator response 
and fiber attenuation properties, models the development of scintillation and
Cherenkov radiation in the active detector materials and simulates 
the light transport, collection, and digitization processes~\cite{response-chain-2015}.
Test stand measurements have been used to adjust the
Birks suppression of scintillation light used in the simulation, and to validate the simulated response 
of the readout electronics~\cite{NOvA-test-stand-2020}.

The ND is located off-axis in the Fermilab Neutrino Main Injector (NuMI) beam 
where it is exposed to a narrow-band $\numu$ flux with a mean energy of 1.86\,GeV.  
The data were taken between August 2014 and February 2017 with the NuMI beam 
operating in the medium-energy, forward horn-current beam configuration.  
The results presented here are obtained from an exposure of 
$8.09 \times 10^{20}$ protons on target (POT).

\section{Simulation of Neutrino Interactions}
\label{sec:Nu-Int-Modeling}
For this analysis, simulation of neutrino events in the ND is based on 
the GENIE v2.12.2 neutrino event generator~\cite{Andreopoulos-NIM-2010, GENIE-2015}.  
This GENIE-based reference Monte Carlo code (MC) has been described in detail  
in a previous publication~\cite{Adjust-Models-2020}. 
In brief, the target nucleus is modeled as a 
local relativistic Fermi gas~\cite{Nieves-PRC-2004} with addition of a high-momentum tail 
for the momentum distribution of single nucleons to account 
for short-range correlations~\cite{Bodek-Ritchie-1981}.
CCQE interactions are simulated 
using weak interaction current--current phenomenology~\cite{Llewellyn-1972}.
Neutrino-induced pion production arises from interactions with single nucleons and 
proceeds either by RES processes 
or by non-resonant shallow and DIS reactions.    
Pion production via RES is simulated using the Rein--Sehgal model~\cite{Rein--Sehgal-1981} 
with incorporation of modern baryon-resonance properties~\cite{PDG-2012}.  
Non-resonant inelastic scattering is modeled 
using the scaling formalism of the Bodek--Yang model~\cite{Bodek-2005-PS} in conjunction 
with a custom hadronization model~\cite{T-Yang-2009} and PYTHIA6~\cite{PYTHIA-6}.
Parameters of DIS processes are
adjusted to reproduce electron and neutrino scattering measurements 
over the invariant hadronic mass range $W<1.7$\,GeV~\cite{Gallagher-2006}.
In particular, a 57\% reduction in the nominal GENIE rate 
for \numu\,CC non-resonant pion production is imposed, as this yields
better agreement with deuterium bubble chamber data~\cite{Wilkinson-PRD-2014, Rodrigues-EurPhys-2016}.
Neutrino-nucleus COH scattering resulting in single pion production is simulated 
using the Rein--Sehgal model~\cite{ReinSehgal:1983, ReinSehgal:2007}.
The reference simulation includes a treatment of final-state intranuclear interactions (FSI) of pions and nucleons that
are created and propagate within the struck nucleus.    An effective model 
for FSI is used in lieu of a full intranuclear cascade;  each pion is allowed to have at most
one rescattering interaction while traversing the nucleus~\cite{Dytman-2011-CP}.  This approximation 
enables event reweighting to be applied to the simulation.

Recent advances in neutrino phenomenology motivate 
additional augmentations to GENIE~\cite{Adjust-Models-2020}.
For CCQE reactions, kinematic distortions attributed to 
screening of electroweak couplings in a nuclear medium are included as
a reweight based on the calculations 
of Nieves and collaborators using the 
random phase approximation (RPA) technique~\cite{Nieves-PRC-2004, RGran-archive}.
For baryon-resonance pion production, 
experiments have reported a suppression effect 
at very low four-momentum transfer, 
$Q^2$~\cite{MiniBooNE-pi0-2011, MiniBooNEPiplus:2011, MINOS-2015, Altinok-2017, Stowell-2019}.
To account for this suppression, a weight analogous to 
 the RPA reweight but parametrized in terms of $Q^2$ instead of $( q_{0}, |\vec{q}\,|)$
 is applied to CC RES events at low $Q^2$, and a systematic uncertainty is assigned to
the RES model.   For $Q^2 \leq$ 0.2 GeV/$c$, the fractional uncertainty on the cross section
associated with RES suppression is $\leq$ 1.5\%.

The analysis uses five different models that describe 2p2h reactions; all of the models are implemented in the GENIE framework.
Three of the models are data-based and two are theoretically motivated.    NOvA tune 2p2h {\it (i)} is a model that has been
adjusted to match the NOvA ND data~\cite{Adjust-Models-2020}.  It was used in previous 
NOvA neutrino-oscillation investigations~\cite{NOvA-Oscillations-2021, NewConstraints-2018, NOvA-PRL-2019},
and it is the 2p2h model used by the reference simulation for this work.    
The other 2p2h models include {\it (ii)} the GENIE Empirical model (or ``Empirical MEC" or ``Dytman MEC")~\cite{Dytman-MEC},
{\it (iii)} a representation of 2p2h designed to match MINERvA inclusive \numu~scattering
data reported in~\cite{Rodriques-2016}, {\it (iv)} the SuSAv2 microscopic MEC model~\cite{Megias:2015, Simo-2016, Dolan-PRD-2020},
and {\it (v)} the microscopic model developed 
by the Val\`{e}ncia group (Nieves {\it et al.}~\cite{Nieves:2011pp, Gran-PRD-2013}).   
In delineating systematic uncertainty associated with 2p2h modeling, re-weighting was
applied to the MINERvA tune~\cite{ddXsec-AQE-MINERvA} and to the SuSAv2 and Val\`{e}ncia models that varied
the relative abundances of final-state hadronic systems consisting of two protons versus a neutron-proton pair.

A main goal of this work is to rate the 
performance of these models in predicting differential cross sections measured using NOvA data.
In figures and tables to follow, 2p2h-model predictions are
displayed in the order enumerated above.    This ordering is highest-to-lowest according to the 
fraction of the CC inclusive cross section ascribed to 2p2h by the reference MC when using each
of the 2p2h models.   Table~\ref{tab:Ratios-2p2h-to-Inclusive} shows the 2p2h fractional contributions predicted for
the NOvA measurement.
\begin{table}
\begin{center}
\caption
{Cross-section ratios $\sigma$(2p2h)/$\sigma_{\text{cc}}$(inclusive) for the NOvA data 
predicted by the reference MC when using, in turn, each of the 2p2h models.}
\smallskip
\label{tab:Ratios-2p2h-to-Inclusive}
\begin{tabular}{cccccc}
\hline
\hline
2p2h Model & Predicted Fraction \tabularnewline
  \small{in Reference MC} &  \small{$\sigma$(2p2h)/$\sigma_{\text{cc}}$(inclusive)}  \tabularnewline
\hline 
NOvA tune 2p2h & 0.27 \tabularnewline
GENIE Empirical & 0.21 \tabularnewline
MINERvA tune 2p2h & 0.17  \tabularnewline 
SuSAv2 2p2h  & 0.13 \tabularnewline
Val\`{e}ncia 2p2h  & 0.11  \tabularnewline
\hline
\hline
\end{tabular}
\end{center}
\end{table}

\section{Event Reconstruction and Selection}
\label{sec:Reco-and-Selection}
Energy deposits (hits) in the detector resulting in APD responses above a noise-vetoing threshold are recorded 
with energy, time, and channel location information.   Calibration of the absolute energy deposition 
of hits is established using intervals of ionization on cosmic ray muon trajectories that enter and range to a stop
in the detector.   Hits neighboring each other in space and time 
are assumed to be associated with a single neutrino interaction.  The hits
are grouped into candidate particle trajectories (tracks) via a Kalman filter-based
algorithm~\cite{Kalman-1960, Ospanov-2008, Raddatz-2016}
in both the horizontal and vertical two-dimensional detector views~\cite{Baird-thesis-2015, NOvA-tracks-2015}.
Tracks from the two views that overlap are combined to form three-dimensional tracks.
A separate algorithm scores the tracks according to a $k$-nearest neighbor classifier~\cite{Altman-1992}  
and assigns the most muon-like track (if one is present) as the muon candidate, using criteria described in the paragraph below.
The track reconstruction examines the most upstream hits of the candidate interaction and determines the interaction vertex
plus emerging line segments that best describe those hits~\cite{Baird-thesis-2015}.
For the hits associated with the reconstructed vertex, 
a different algorithm is used to form particle trajectories (prongs)~\cite{Niner-thesis-2015}.
The latter algorithm allows hits to be more broadly distributed around the particle's direction, 
and it is optimized for electromagnetic shower reconstruction.

Candidate \numu\,CC interactions are selected using procedures previously developed for the NOvA measurement
of the CC inclusive double-differential cross section in muon kinetic energy, $T_{\mu}$, and muon production angle, 
$\cos \theta_{\mu}$~\cite{NOvA-CC-inclusive}.   The signal definition is the 
same as in that previous work.   Specifically, a signal event is a true $\numu$ CC event whose primary interaction
vertex lies within the fiducial volume.  Furthermore a signal event must satisfy the the muon kinematics selections
stated in the next paragraph below as expressed in true values of $T_{\mu}$ and $\cos \theta_{\mu}$. 
Included as signal is a tiny subset of true $\numu$ CC events (estimated 7 of 995,760 events) for which 
the reconstructed muon candidate is actually a charged pion rather than the final-state muon.

 Events that pass basic quality cuts in timing, containment, and 
contiguity are required to have a candidate muon track.   
Muon identification is based on a multivariate algorithm that examines
hit-to-hit energy deposition and multiple scattering.   
Muons are distinguished from charged pions on the basis of 
{\it (i)} the difference between log-likelihood functions based on $dE/dx$ of muons versus pions,
{\it (ii)} average $dE/dx$ in hits in the last 10\,cm of tracks, {\it (iii)}  average $dE/dx$ in hits 
in the last 40\,cm of tracks, and {\it (iv)}  muon versus pion likelihood 
assigned according to average angular deflections as a function of distance traveled.    
These reconstructed variables are processed using a boosted decision tree algorithm~\cite{NOvA-CC-inclusive}.
The event vertex is placed at the beginning of the muon track, 
and it is required to lie within a fiducial volume of dimensions 2.7\,m by 2.7\,m by 9.0\,m that is contained within
the detector's active volume.  The fiducial volume begins one meter downstream from the front face
of the active volume and is surrounded on all sides by at least 52\,cm of active volume.
To ensure reliable estimation of final-state
hadronic energy, events having hit clusters that extend to the edges of the ND are rejected.
Furthermore, events are rejected if any track or prong other than the 
muon enters the muon catcher.   The energy of muons that stop in the detector, 
$E_{\mu}$, is determined using track length.  The energy resolution is 4\% for muons that stop in the ND
scintillator volume upstream of the muon catcher, while for muons that stop in the catcher it is typically 5\% to 6\%~\cite{NOvA-CC-inclusive}.

In order use the tracking volume of the NOvA ND to carry out a \numu\,CC inclusive measurement in an optimal way,
the analysis imposes requirements on final-state muon kinematics.  These requirements, as described below, are 
the same as were used previously for the NOvA measurement of 
$d^2\sigma_{\rm{incl}}/d \cos\theta_{\mu} dT_{\mu}$~\cite{NOvA-CC-inclusive}.
The requirements are applied to the signal definition and to selection cuts on reconstructed events.   
They have an impact on the shape of the extracted cross section.   For the signal definition the
requirements are defined in terms of eight intervals in true $T_{\mu}$, each of which is paired with
an interval in $\cos \theta_{\mu}$.    A summary of the allowed pairs of ranges in 
is given in Table~\ref{tab:Muon-Kin-Bins} where, for example, selected muons with $T_{\mu}$ between 0.5--1.1 GeV 
must have $\cos \theta_{\mu}$ values within 0.5--1.0, and similarly for the remaining pairs of ranges in the Table.

\begin{table}
\caption{\label{tab:Muon-Kin-Bins}  Muon kinematic requirements of the signal definition for this
analysis.   Selected muons have ($T_{\mu}$, $\cos \theta_{\mu}$) values that fall within
the eight pairs of  intervals delimited by the vertical columns of the Table.}
\centering{}
\begin{tabular}{cc|c|c|c|c|c|c|c}
\hline 
\hline
$T_{\mu}$(GeV)\,0.5 to: & 1.1 & 1.2 & 1.3 & 1.4 & 1.8 & 1.9  & 2.2  &  2.5    \tabularnewline
\hline 
$ 1.0\, \ge\cos \theta_{\mu} \ge$:& 0.5 & 0.56 & 0.62 & 0.68 & 0.85 & 0.88   &  0.91  & 0.94    \tabularnewline
\hline
\hline 
\end{tabular}
\end{table}

Selected events are binned and unfolded using \qthree\,
and \eavail\, and no cuts are imposed using these variables. 
The analysis is restricted to the kinematic domain 0.0 GeV/$c$ $\leq$\,\qthree\,$\leq$\,2.0\,GeV/$c$ and \eavail\,$\leq$\,2.0
GeV.  Regions with larger values of \qthree\, and/or \eavail\, have negligible event statistics.
With final results, bins with cross-section total uncertainty exceeding 100\% are not reported.   This criterion removes
five bins (visible in Fig.~\ref{Fig08}) located along the diagonal kinematic boundary that defines highest \eavail\, per
interval in \qthree .

\section{Variables, Binning, and Cross Section}
The observables used to construct other analysis variables
are the muon energy, $E_{\mu}$, the muon momentum, $p_{\mu}$, 
the muon angle with respect to the neutrino beam direction,
$\theta_{\mu}$, and the sum of the calibrated, observed (visible) hadronic energy deposited
in the detector, $E_{\rm{vis}}$.   The thresholds for reconstructing charged pions and protons as
tracks are $T_{\pi} \simeq$ 200\,MeV~\cite{NOvA-low-Ehad} and $T_{\text{p}} \simeq$ 275\,MeV respectively, 
with tracking efficiencies at thresholds being $\simeq$ 35\%.
However, charged hadrons of few tens of MeV kinetic energy can produce detectable 
ionizations in the NOvA ND.   The fully reconstructed energy 
of the final state hadronic system, $E_{\rm{had}}$, is
obtained by applying correction weights to $E_{\rm{vis}}$ that account for unseen energy, such
as that lost to inactive detector material or carried away by neutrons.  
The energy resolution for $E_{\rm{had}}$ in this analysis is 30\%~\cite{Psihas-thesis-2018}.   
The reconstructed neutrino energy, $E_{\nu}$, is calculated as the 
sum of $E_{\rm{had}}$ and $E_{\mu}$.
The four-momentum-transfer-squared, $Q^{2}$, from the leptonic current 
to the hadronic system is calculated as 
\begin{equation}
\label{def-Q2}
Q^2 = -(k-k')^2 =  2E_\nu(E_{\mu}- p_\mu\cos\theta_{\mu})-m_\mu^2.
\end{equation}
As previously noted, theoretical treatments of 2p2h are often couched in terms of
magnitudes of four-momentum-transfer components, 
$q_{0}$ and $|\vec{q}\,|$.    The two variables on which this analysis
is based are ones that approximate these components;  their construction
is described below.

\smallskip
 \noindent
{\it Three-momentum transfer}:
The magnitude of the three-momentum transfer, $|\vec{q}\, |$, from the leptonic current 
to the target nucleus is calculated as follows:
\begin{equation}
\label{eq:Vecq}
|\vec{q}\, | = \sqrt{Q^{2}+(E_{\nu}-E_{\mu})^{2}}~.
\end{equation}
The relationship between reconstructed
and true \qthree\, is established using selected events from the reference
simulation.   It is linear to good approximation, and the variance from linear
is measured by the absolute resolution for reconstructed \qthree, defined
as the standard deviation, $\sigma$, of the distribution of the absolute residual,
(\qthree$_{\text{true}}$ - \qthree$_{\text{reco}}$).
The absolute resolution for \qthree\, is 0.28 GeV/$c$.
The fractional \qthree\, resolution is similarly defined
as $\sigma$ of the fractional \qthree\,
residual distribution; it is 21\%. 
The distributions of absolute and fractional residuals broaden with increasing \qthree\,, however
they remain centered very close to 0~\cite{Olson-thesis-2021}.     

\smallskip
\noindent
{\it Available energy}:
A second variable is needed to characterize 
the energy transfer received by the hadronic system. 
The variable $E_{\rm{avail}}$ (see Sec.\,\ref{sec:Intro}) is designed 
to be as close as possible to the energy that can be reliably
observed in the detector with minimal model dependence.   
Available energy is constructed by correcting 
$E_{\rm{vis}}$ to the amount of visible energy that would be detected 
in a perfect detector.
 
Reconstruction of \eavail\, is based on a map from event 
visible energy, $E_{\rm{vis}}$, to true \eavail\,, constructed using selected 
MC events.   For each event, the 
sum of reconstructed non-leptonic energy deposited in the detector, 
$E_{\rm{vis}}$, is matched with the true \eavail\, value.   
Then, for each bin (width = 20\,MeV) of reconstructed $E_{\rm{vis}}$, the mode of the true \eavail\, distribution
is obtained.  A profile of the modes is then fitted to a function that transforms reconstructed $E_{\rm{vis}}$ to true \eavail\,~\cite{Olson-thesis-2021}.
Various polynomial functions were tried, with variable coefficients allowed for successive integer powers of $E_{\rm{vis}}$.
The mapping to modes of true \eavail\, from reconstructed $E_{\rm{vis}}$ is in fact nearly linear; no improvement of fit $\chi^2$/DoF is observed
with inclusion of cubic or higher-power terms.  Consequently a quadratic form suffices to describe the
relationship: $E_{\rm{avail}}=a + b\, (E_{\rm{vis}})+c\, (E_{\rm{vis}})^{2}$. 
The linear term has slope $b = 1.68\pm 0.03$;  it requires a 
quadratic correction ($c = 0.024\pm0.008\, \text{GeV}^{-1}$) and a small offset ($a = -0.0051\pm 0.0032$\,GeV).

 The absolute  \eavail\, resolution,
defined as $\sigma$ of the absolute residual distribution, is 0.21\,GeV. 
The fractional  \eavail\, resolution is 32\%.   The \eavail\, residual distributions,
when broken out into bins of increasing \eavail, remain centered near zero
with approximately Gaussian shapes that broaden with bin
energy~\cite{Olson-thesis-2021}.   

\noindent
{\it Resolution binning}:
Bins of variable width are chosen for each of the two kinematic variables according to the
experimental resolutions~\cite{Olson-thesis-2021}.  To cover 
the interval $0 \leq$\,\qthree\,$\leq 2.0$\,GeV/$c$, twelve bins are chosen
whose widths become larger with increasing \qthree.  
An overflow bin is allotted for the few events that have \qthree\, $>\,2.0$\,GeV/$c$.
Similarly for \eavail, since the resolution worsens with increasing values in a linear way
over the range from 0 to 2.0\,GeV, nine bins with increasing
widths are chosen to span this interval (together with an overflow bin).   With these choices,
each bin contains more than 1000 events.   The net result is the 2-D pixelation of the \qthree\,-\,\eavail~phase space
that is apparent in figures to follow.

\smallskip
\noindent
{\it Double-differential cross section}:
The flux-integrated double-differential cross section is calculated
as follows:
\begin{equation}
\left(\frac{d\sigma^{2}}{d|\vec{q}|\,dE_{\rm{avail}}}\right)_{ij}
= \frac{\sum_{\alpha\beta}U_{ij,\alpha\beta}\left(N_{\alpha\beta}^{\rm{Sel}}-N_{\alpha\beta}^{\rm{Bkgd}}\right)}
{\epsilon_{ij}\,\Phi_{\nu}\,T_{N}\,(\Delta|\vec{q}|)_{i}\,(\Delta E_{\rm{avail}})_{j}}.
\label{eq:XSec}
\end{equation}
\noindent
The array $N_{\alpha\beta}^{\rm{Sel}}$ is the number of selected data
events, and $N_{\alpha\beta}^{\rm{Bkgd}}$ is the number of estimated background
events that is subtracted from the data to get the estimated signal.
The unfolding matrix, $U_{ij,\alpha\beta}$, converts event counts in reconstructed
bins ($\alpha,\beta$) to counts in unfolded bins $(i,j)$; $\epsilon_{ij}$ is
the efficiency correction in the (\qthree,\eavail) bin designated
by $(i,j)$, $\Phi_{\nu}$ is the integrated neutrino flux, $T_{N}$
is the number of nucleons in the fiducial volume, and $(\Delta$\qthree$)_{i}$
and $(\Delta E_{\rm{avail}})_{j}$ are the widths of the bin $(i,j)$.

\begin{figure}
\begin{centering}
\includegraphics[scale=0.43]{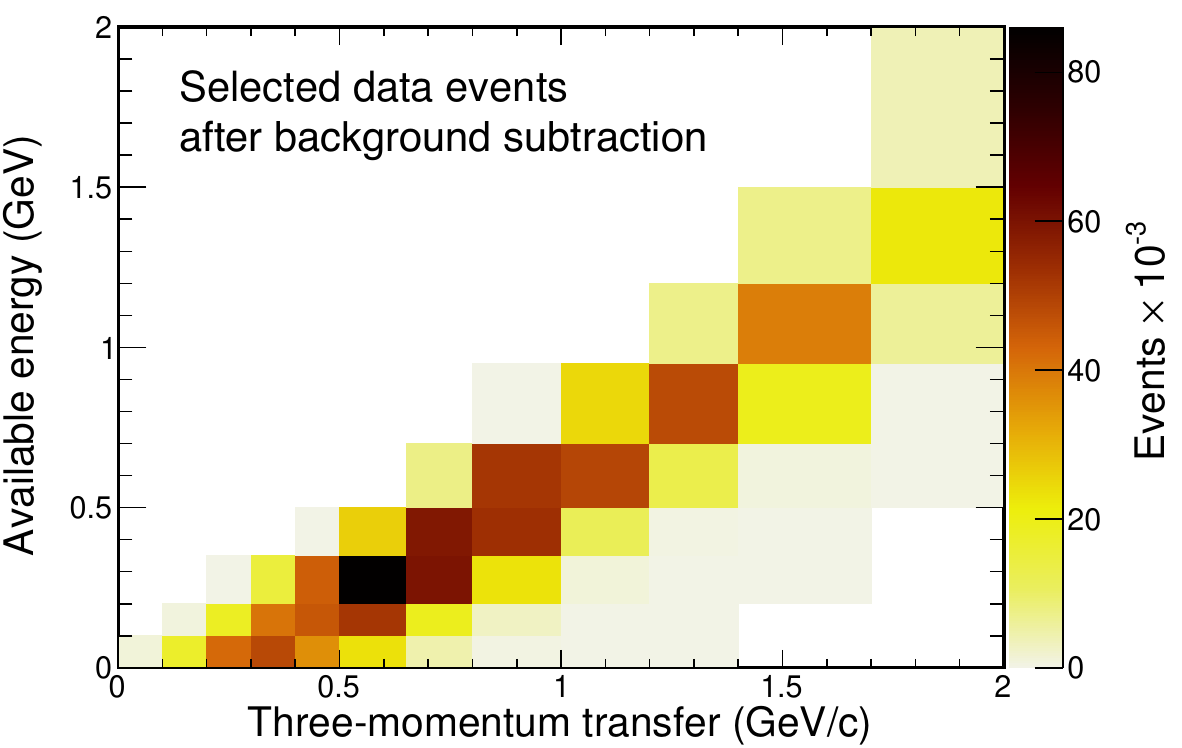}
\par\end{centering}
\caption{\label{Fig02} Distribution of selected signal events of the data, after background subtraction.}
 \end{figure}

\section{Selected Sample}
\label{sec:Selected}
The selected data sample consists of events that reconstruction indicates have occurred 
in the kinematic domain $0 \leq$ \qthree\,$\leq$ 2.0 GeV/$c$ and $0 \leq$ \eavail\,$\leq$ 2.0 GeV.
The inclusive CC signal-event data sample is obtained
by subtracting the estimated background from the selected data (see Sec.\ref{sec:Backgrounds}).
The signal-event sample consists of 995,760 events whose distribution over the \qthree\,-versus-\eavail\, plane is
shown in Fig.~\ref{Fig02}.

\begin{figure}
\begin{centering}
\includegraphics[scale=0.43]{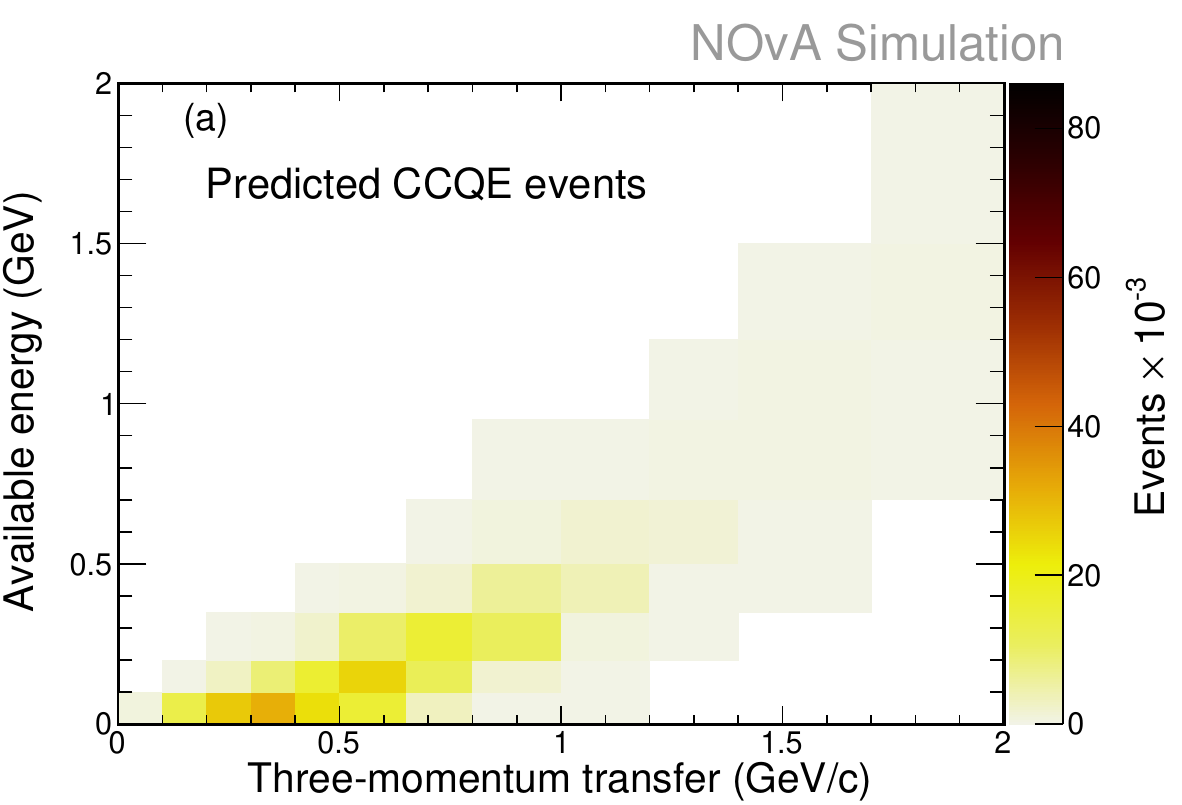}
\includegraphics[scale=0.43]{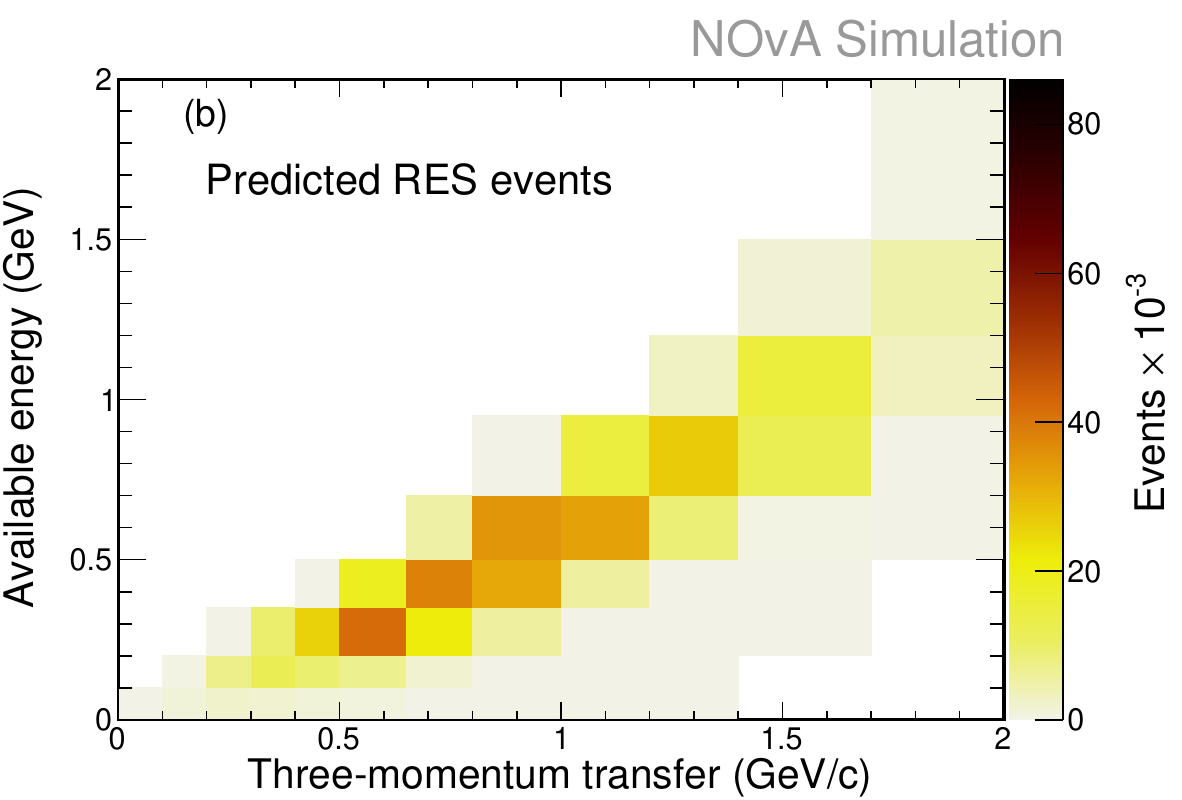}
\includegraphics[scale=0.43]{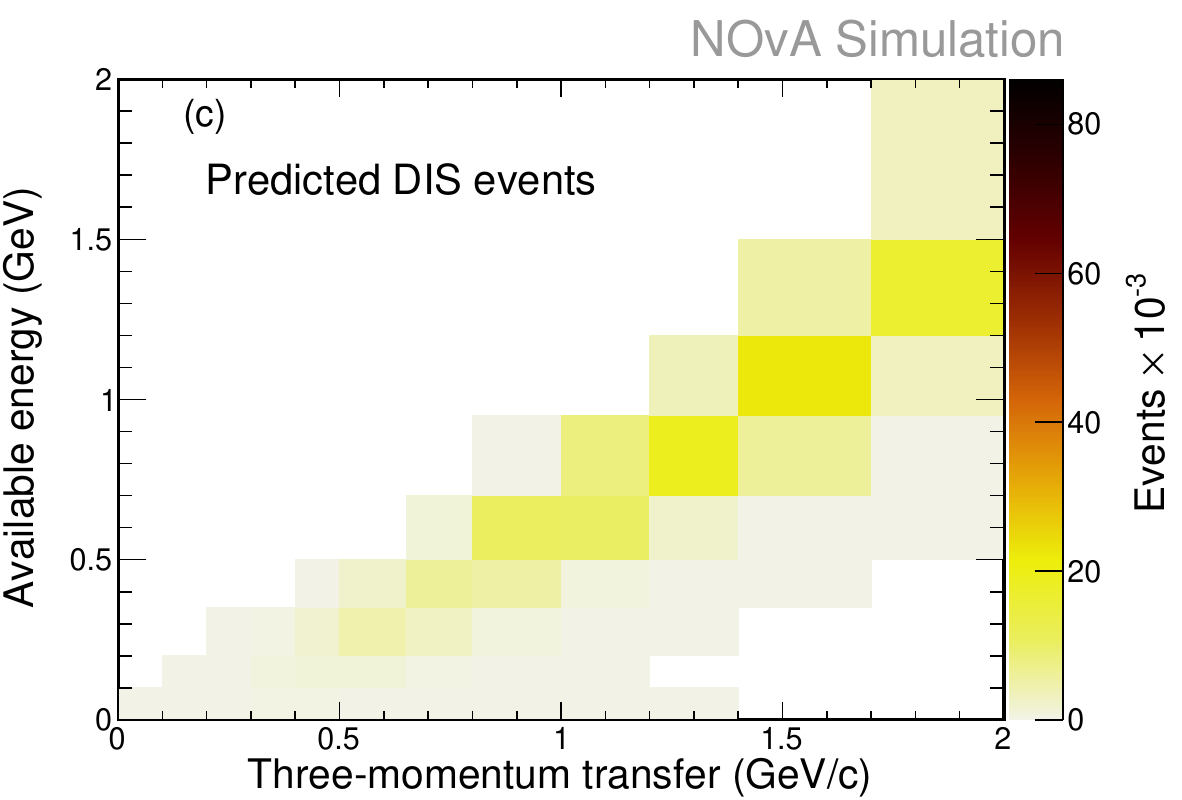}
\par\end{centering}
\caption{\label{Fig03} Event distributions predicted for the data exposure from CC interactions on single nucleons 
of the NOvA nuclear medium from the reaction channels CCQE (a), RES (b),
and DIS (c).}
\end{figure}

The majority of the data events that populate Fig.~\ref{Fig02} are predicted to arise from the known CC neutrino-nucleon 
interactions CCQE, RES, and DIS that occur within the NOvA nuclear medium.  
Plots (a), (b), and (c) of Fig.~\ref{Fig03} display the separate event distributions predicted for the three CC channels,
for the same POT exposure and with the same selections as applied to the data of Fig.~\ref{Fig02}.
The three channels differ significantly in their absolute rates and in the locations of their peak event rates.
The CCQE interactions dominate the region of low \qthree\, and low \eavail\, where the 2p2h process is also 
expected to have a sizable presence.   The distribution for RES reactions overlaps portions of the CCQE
region, however it is most abundant in regions with \qthree\,$\geq 0.5$\,GeV/$c$ with \eavail\,$\geq 0.2$\,GeV.    
Above \qthree\,$\simeq 1.2$\,GeV/$c$ with \eavail\,$\ge 0.7$\,GeV, the RES distribution 
drops off while the DIS distribution gains strength.  The DIS contribution is largest in the vicinity
of \qthree\,$\simeq 1.5$\,GeV/$c$ and \eavail\,$\simeq 1.0$\,GeV.

\section{Sample Efficiency and Purity}
\label{sec:Efficiency-Purity}
\begin{figure}
\begin{centering}
\includegraphics[scale=0.43]{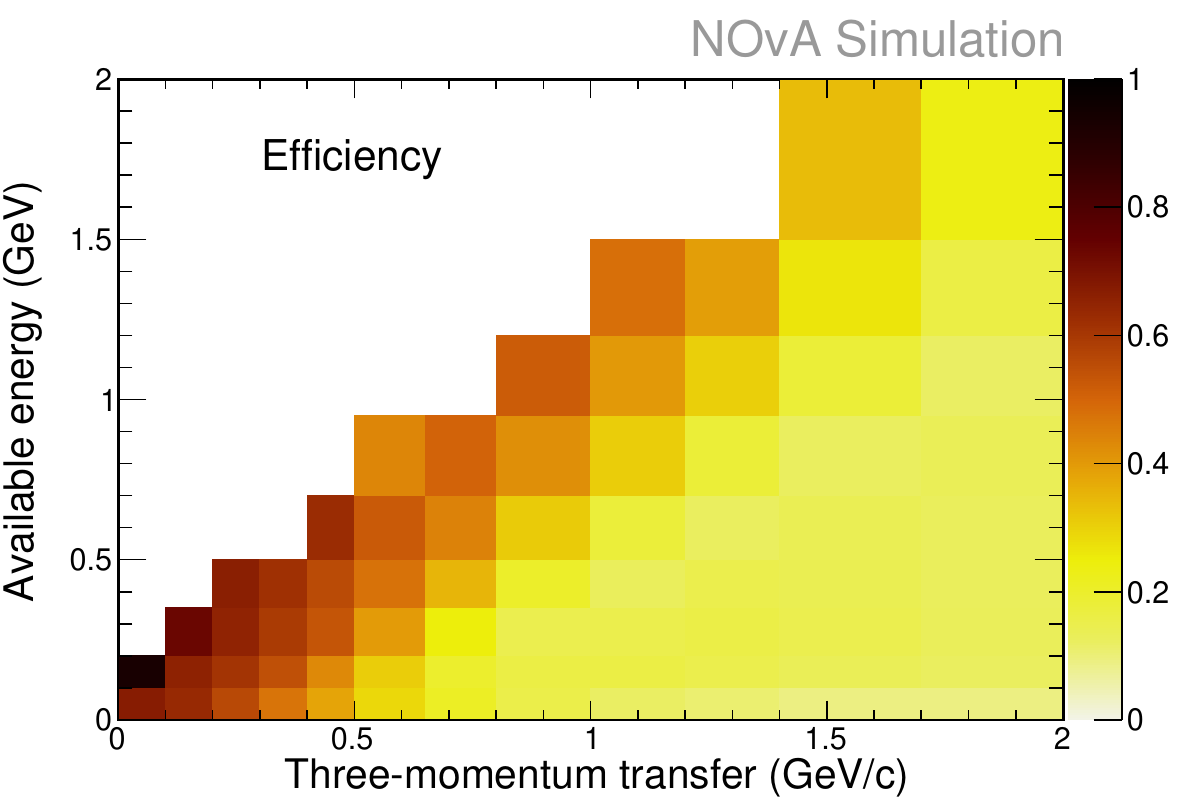}
\includegraphics[scale=0.44]{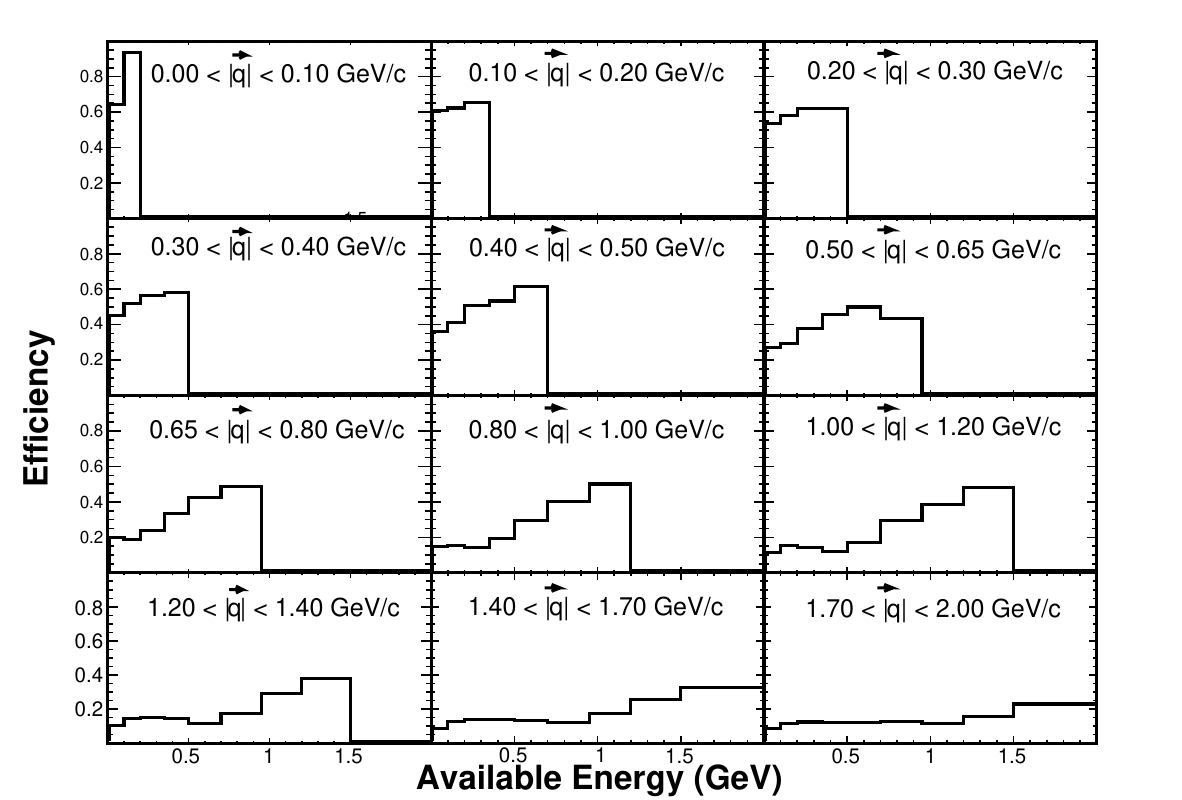}
\par\end{centering}
\caption{Companion displays of event selection efficiency.  The upper diplot shows the variation
in efficiency over the \qthree\,-vs-\eavail\, plane.  The panels of the lower plot show efficiency versus
\eavail\, for nine intervals in \qthree.   The efficiency peaks along the kinematic boundary
and diminishes smoothly with increasing displacement from the boundary.}
\label{Fig04}
\end{figure}

\begin{figure}
\begin{centering}
\includegraphics[scale=0.43]{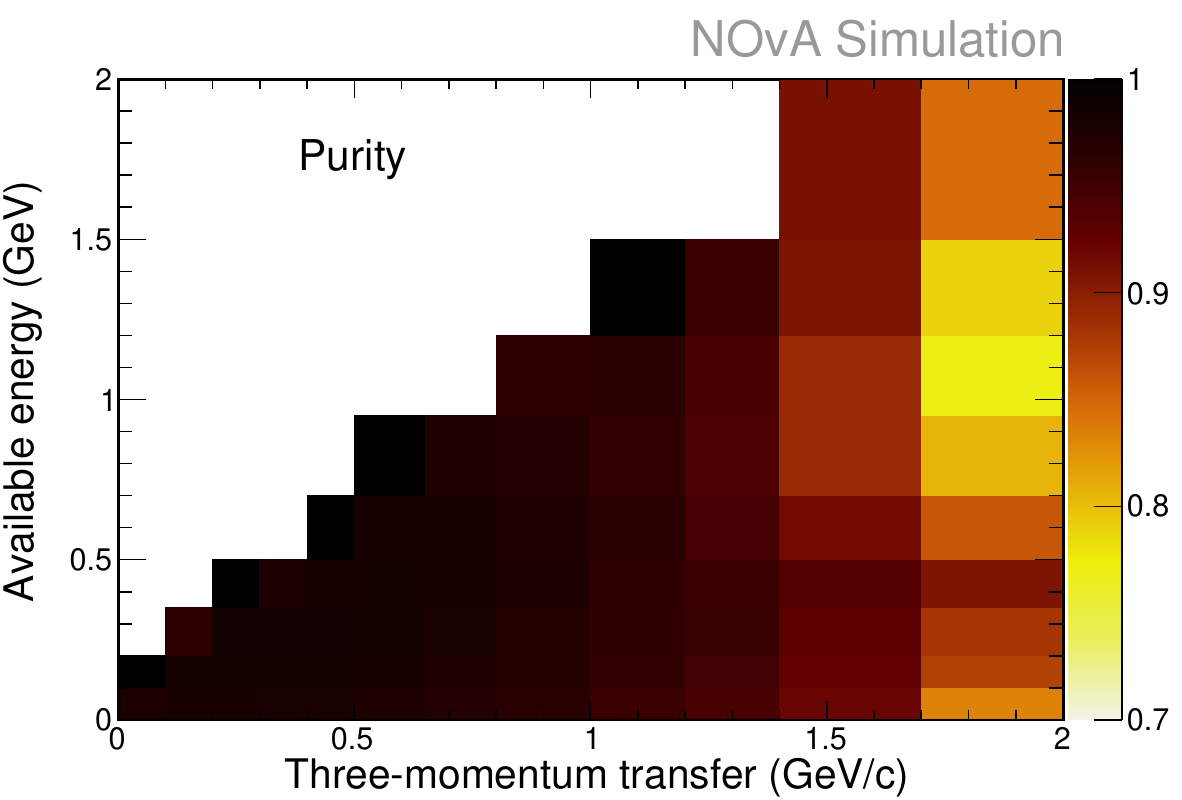}
\par\end{centering}
\caption{The sample purity in bins of (\qthree\,, \eavail\,). 
The purity is fairly uniform over the analyzed phase space, with an
average value of nearly 93\%.}
\label{Fig05}
\end{figure}

The cross section requires the correction factors, $\epsilon_{ij}$,
for sample selection efficiency, defined as the fraction of true signal events 
that are selected according to the signal definition of Sec.~\ref{sec:Reco-and-Selection}.
Also required is bin-by-bin knowledge
of the selected sample purity, i.e., the fraction of signal events among selected
events, in order to implement the subtraction of background 
from the selected sample (Sec.\,\ref{sec:Backgrounds}).
Figures~\ref{Fig04} and~\ref{Fig05} show the selection efficiency and purity, respectively,
over the (\qthree\,, \eavail\,) kinematic plane.    

The requirements of full containment for muon tracks and 
for final-state hadrons have a major impact on the selected sample. 
Their effect gives rise to correlations in detection efficiency 
between \qthree\, and \eavail\, that can be seen in the companion
diplot and projection displays of Fig.~\ref{Fig04}.   
Regions of high \qthree\, with low to intermediate \eavail\, 
(areas towards lower right in Fig.~\ref{Fig04}(top), left-most bins
of lower projections in Fig.~\ref{Fig04}(bottom))
have a relatively lower detection efficiency.
The efficiency is highest (40\% to nearly 100\%)
along the kinematic boundary where the final-state energy is roughly balanced between the 
leptonic and hadronic systems.   In regions remote from the boundary, the CC interactions tend 
to have higher momentum (i.e., longer muon tracks) and these have a lower probability of stopping within the fiducial volume.
Consequently the efficiency falls off smoothly and rather rapidly with increasing displacement from the kinematic edge.
The region with \eavail\,$<0.4\,$GeV and $0.6\leq$\,\qthree$\,\leq1.2\,$GeV/$c$
has a slowly varying selection efficiency that averages around 20\%.
Quasielastic scattering and multi-nucleon scattering occur predominantly in lower regions of \qthree\, and \eavail, while
baryon-resonance production and deep inelastic scattering dominate higher \qthree\, and \eavail.

The average efficiency for the selected sample reflects 
the cost of the selection cuts that are required to minimize
background contributions.   Starting from a raw sample of selected CC events, the 
muon identification cut gives an event reduction of nearly 15\%, and 
muon containment plus vertex containment give an additional reduction of nearly 53\%.   
Subsequent restrictions on the allowed muon phase space and on hadronic shower containment
give an additional 4.5\% reduction, resulting in a final average efficiency of 27.8\%.


Figure~\ref{Fig05} shows sample purity in bins of reconstructed available energy versus
 three-momentum transfer with binning determined by the resolution.
The purity after all selections is fairly 
uniform across the analyzed phase space and it exceeds 75\% in nearly all bins.
The average purity over all bins is 92.9\%.  
In contrast to efficiency, the purity exhibits relatively mild correlations 
between \qthree\, and \eavail.   The purity is highest in bins wherein
 \eavail\, in GeV is roughly equal to the value of \qthree\, in GeV/$c$; it is diminished by 10\% to 15\%
in regions where \qthree\, is numerically 
larger than \eavail\, or where \qthree\, exceeds 1.7 GeV/$c$.  As described below, two of the four types of
background reactions tend to appear in those regions.

\section{Background processes and data unfolding}
\label{sec:Backgrounds}
The selected data events include a 7.2\% contribution from background events 
as estimated using the reference simulation.
Nearly all background events fall into one of four categories:
{\it (i)} \anumu\, interactions arising
from the defocused component of the \numi\, beam (2.8\%), 
{\it (ii)} \numu\,CC events whose true muon kinematics (but not their reconstructed kinematics)
 fail the $T_{\mu}$ and/or $\cos \theta_{\mu}$ selections (2.3\%),
{\it (iii)} NC interactions or \nue-flavor CC events reconstructed
as \numu\,CC interactions (1.2\%), and {\it (iv)} \numu\,CC events with vertices originating outside 
the fiducial volume (including events with an interaction in the rock) (0.9\%).
The background processes distribute over the analyzed (\qthree\,, \eavail ) phase space and 
their subtraction does not significantly change the
shape of the signal distribution.   There is however a mild
tendency for processes {\it (ii)} and {\it (iii)} to appear in regions with \qthree\, $>$ 1.0\,GeV/$c$ and
with \eavail\, $>$ 0.5\,GeV.

The distribution of events in bins of reconstructed \qthree\, and \eavail\, is subject to distortions
induced by finite detector resolution.   This detector-induced smearing is corrected by subjecting
the event distribution to the D'Agostini iterative unfolding algorithm~\cite{Schmitt-2017,D'Agostini-NIM-1995}
as implemented by the RooUnfold package in ROOT~\cite{RooUnfold}.  
The `truth' that underwrites the data is of course unknown, consequently the optimal number of iterations
to use in unfolding the data was evaluated using 500 independent,
systematically shifted simulation samples (referred to as ``universes").  Each universe is constructed 
by randomly altering all of the physics parameters used by the GENIE event generator.   The randomization
of each parameter is based on a gaussian distribution about the nominal value whose width 
is the 1-$\sigma$ uncertainty assigned to that parameter.

The evaluation of unfolding iterations was based on ensemble averages of the mean squared error~\cite{Schmitt-2017}.
The mean squared error per universe (MSE) is defined as
\begin{equation}
\label{eq:MSE}
\text{MSE}\,=\,\sum_{j=1}^{\text{Bins}}\frac{(\sigma_{\text{Unfold}_{j}})^{2} + (\text{Unfold}_{j}-\text{True}_{j})^{2}}{(\text{True}_{j})^2}.
\end{equation}
Here, $\sigma_{\text{Unfold}_{j}}$ is the error assigned by RooUnfold to the $j$th bin,
$\text{Unfold}_{j}$ is the event count in the $j$th bin of the unfolded
distribution, and $\text{True}_{j}$ is that in the $j$th bin in the truth distribution.
The MSE value averaged over the ensemble 
of 500 universes, $\overline{\text{MSE}}$, was used
to determine the best number of unfolding iterations.
The $\overline{\text{MSE}}$ was observed
to reduce dramatically with one iteration and to minimize with two iterations; with 
iterations beyond two it gradually and continuously climbed away from the minimum.   The same behavior
was observed using a $\chi^{2}$ constructed as 
$\sum_{j=1}^{\rm{Bins}}\frac{(\rm{Unfold}_{j}-\rm{True}_{j})^{2}}{\sigma_{\rm{Unfold}_{j}}}$.  Additionally, 
the ratios of unfolded distributions to MC truth distributions over the \qthree\,-vs.-\eavail\, plane were
examined using fake data studies, and the results indicated two unfoldings to be optimal.  For these 
reasons the analysis proceeded with two unfolding iterations applied to the data~\cite{Olson-thesis-2021}.
The verity of this procedure was then checked by examining the $\chi^2/DoF$ over all bins for successive
unfoldings of the data, compared to the data distribution obtained with two unfoldings.   With this metric the data
exhibits behavior very similar to that observed in the simulation studies. That is, two unfoldings coincides with
minimum $\chi^2$, while six more unfoldings yield larger $\chi^2$ values than does a single unfolding.

\section{Systematic Uncertainties}
\label{Sec:Systematics}
The cross-section measurement requires knowledge of neutrino-nucleus interactions
including 2p2h, of the neutrino flux, of detector calibration and response, and of 
ionization and Cherenkov light initiated by final-state particles.  There are uncertainties associated with 
each of these quantities.  Most of the sources of uncertainty that affect the present work
were encountered in the NOvA measurement of CC inclusive $d^{2}\sigma/d\cos\theta_{\mu}\,dT_{\mu}$
and details of their treatment are given in Ref.~\cite{NOvA-CC-inclusive}.
As in the previous work, this analysis uses the multi-universe method for determining the total 
systematic uncertainty.  The method involves randomly
varying parameters that characterize uncertainty sources to create a new prediction
 -- a ``universe."  In the new simulation the background estimate is altered,
as are the unfolding matrix, efficiency correction, and flux estimation
for the cross-section calculation.   Consequently the new simulation
leads to a variant cross section for this particular universe.  
The ensemble of variant cross
sections is then compared
to the reference simulation used by the analysis.   The error band is constructed
by taking the root mean square of the bin-by-bin upward excursions and, separately, the downward excursions.
The resulting error band may, in general, be asymmetric.

The statistical uncertainty arising from the data is ascertained by Poisson-fluctuating the bin contents of the selected data
and then carrying out the entire analysis.  Similarly, the MC statistical uncertainty is accounted for by
Poisson-fluctuating the MC-truth bin contents for a given universe.   Events of that 
fluctuated sample, with subsequent reconstruction and selections, are used in 
calculating the background subtraction, the unfolding response matrix, and the efficiency correction.
These procedures are carried out for 10$^5$
independent trials, and the outcomes are used to calculate the contribution 
to the covariance matrix arising from finite event statistics.

\subsection{Sources of uncertainty}
\label{subsec:sources-of-uncertainty}
There are 96 individual parameters that characterize sources of 
uncertainty; however each of them can be assigned to one of the following 
four general categories:

\smallskip
\noindent
{\it Neutrino flux modeling}:  Sources of uncertainty associated 
with calculation of the forward horn-current neutrino
flux (see Sec.~\ref{sec:ND-nf-de})
include focusing of the primary proton beam, modeling of hadron production in the target and
of secondary production in the horns, and modeling of the beam optics, including
uncertainties in the locations of beamline elements~\cite{NOvA-CC-inclusive}.  The flux uncertainty
acts predominantly as a normalization uncertainty that can introduce high correlations among data bins.

\smallskip
\noindent
{\it Neutrino-nucleus interaction modeling}:  
The reference simulation is used to estimate backgrounds, correct for efficiency
losses, and construct the unfolding matrix. 
Consequently uncertainties
in the parameters of the GENIE-based cross-section modeling propagate 
to the error band of the measurement.  The parameters are those associated with
neutrino-nucleus cross sections, modeling of nuclear effects,
hadronization in neutrino final states, and intranuclear propagation and
scattering of mesons and nucleons~\cite{GENIE-2015, Adjust-Models-2020}.
The modeling of 2p2h interactions receives special treatment, 
as detailed in Sec.\ref{subsec:2p2h-systematic} below.

\smallskip
\noindent
{\it Energy scale}:
The muon energy is measured from its range.
Uncertainties in the muon energy scale arise from modeling the $dE/dx$ energy loss in propagation
through the active scintillator volume and in the muon catcher.
The uncertainty on muon energy includes a component that is uncorrelated between these two
regions of the detector.  
There are uncertainties in the energy scale for ionizations by protons and charged pions.
Visible hadronic energy is used to estimate \eavail\, and the small energy depositions
that arise from secondary neutron scattering affect the estimate. 
The detector's response to neutrons is assigned an uncertainty based upon 
data versus MC comparisons of neutron-enriched samples 
induced by antineutrino QE-like scattering~\cite{NOvA-CC-inclusive}.

\smallskip
\noindent
{\it Detector response}:
There are uncertainties associated with the calibration of the visible hadronic energy scale
and with modeling of the transport of
light produced in the scintillator and wavelength-shifting fibers to the APDs.
The calibrated energy response varies with distance from the readout, and there is uncertainty
in the modeling of its non-uniformity which is
included as a calibration uncertainty.
There are uncertainties in the amount of scintillator light expected
from particles, including that associated with the parameter
of Birks' empirical formula~\cite{Birks}. 
The latter uncertainties are constrained by measurements of the light yield
from protons carried out using a test stand~\cite{Scintillator-test-bench}.  
Light production in the scintillator includes Cherenkov light,
for which there are modeling uncertainties as well.  
Light calibration uncertainties are estimated based on dedicated MC simulations. 
These uncertainties are not included in the multi-universe approach; instead, a covariance matrix is
calculated for each calibration systematic, and these are added to the multi-universe covariances.

In the reference simulation, secondary interactions of produced hadrons with the detector medium are
modeled using Geant4 and there are uncertainties associated with the hadron-nucleus cross sections
that are utilized by the Geant4 code.
 The effect of uncertainties from secondary hadronic scattering was examined 
using simulations wherein the rate of secondary interactions in selected events was
enhanced or diminished by up to 30\%, with the total number of events held constant. 
These changes, when propagated to determinations of the differential cross sections in this
work, generate fractional uncertainties at the sub-percent level for all bins.

The number of nucleons in the fiducial volume is 
$(5.689 \pm 0.014)\times10^{31}$, which gives a negligible contribution 
to the total uncertainty budget.

Sources of uncertainty worthy of note as sizable but likely amenable to reductions in the future, are as follows:
{\it (i)} modeling of the neutrino flux (as discussed in the first paragraph of this subsection) ,
{\it (ii)} modeling of 2p2h processes (discussed in Sec.~\ref{subsec:2p2h-systematic}),
{\it (iii)} the mass parameter of the axial vector dipole form factor in
 CC baryon-resonance production,
 {\it (iv)} the mass parameter of the axial vector dipole form factor in
 CC quasielastic scattering,
 {\it (v)} the shape of RPA enhancements in \qthree\, and
$q_{0}\,$ distributions, and
 {\it (vi)} the mass parameter of the vector form factor in
 CC baryon-resonance production.
 A quantitative breakdown of the total systematic error budget 
 is presented in Sec.~\ref{subsec:total-systematic}.

\subsection{Systematic for 2p2h modeling}
\label{subsec:2p2h-systematic}
The analysis incurs uncertainties from the modeling of
2p2h, reflecting the current limited knowledge about these processes.
To determine the cross-section variations that 2p2h uncertainties
may introduce, the five 2p2h models identified
in Sec.~\ref{sec:Nu-Int-Modeling} were investigated~\cite{Olson-thesis-2021}.
All of the models have a similar cross-section dependence on \enu, but they predict different
absolute rates and distributions over the plane of \qthree\, and \eavail. 
In general, the data tunes give higher event rates over much of the phase space.
The Val\`{e}ncia and SuSAv2 models predict peaks at slightly
higher values of \qthree\, (0.8 GeV/$c$ versus 0.6 GeV/$c$) than do the data tunes 
(see the upper plot of Fig.~\ref{Fig10}).

The shapes of the predicted 2p2h distributions are influenced by the initial state 
dinucleon fraction $R_{N}$ = $(np\rightarrow pp)/(nn\rightarrow np)$ used by the models.
The MINERvA data tune offers a base model in which $R_{N} = 2.8$, together with
two companion tunes in which the final state dinucleon is only $pp$ or $np$.
The GENIE Empirical model uses $R_{N} = 4.0$.  The Val\`{e}ncia and 
the SuSAv2 models each use their own calculated prediction for the di-nucleon fraction.
In the Val\`{e}ncia model $R_{N} = 2.8$, while in the SuSAv2 model $R_{N} = 7.8$.

The MINERvA tune to Val\`{e}ncia 2p2h, which is a data-driven construction, and the SuSAv2
model, as a developed theoretical model, offer 
predictions about 2p2h that are entirely free of tuning to NOvA data.
Moreover the differences between their predictions roughly span 
the variability that occurs among all five of
the models examined~\cite{Olson-thesis-2021}.
Consequently, on a bin-by-bin basis, the largest
excursion from nominal (based on the NOvA tune) predicted by either the MINERvA tune with 
$R_{N} = 2.8$ or by SuSAv2 is taken as the estimate of the uncertainty. 
That is, the largest absolute deviation from the nominal, either positive or negative,
is used to define an error that is symmetric about the nominal.
This 2p2h modeling uncertainty is added in quadrature,
bin-by-bin, with the other sources of systematic error to get the total
systematic uncertainty.   

\subsection{Total systematic uncertainty}
\label{subsec:total-systematic}
\begin{figure}
\begin{centering}
\includegraphics[scale=0.47]{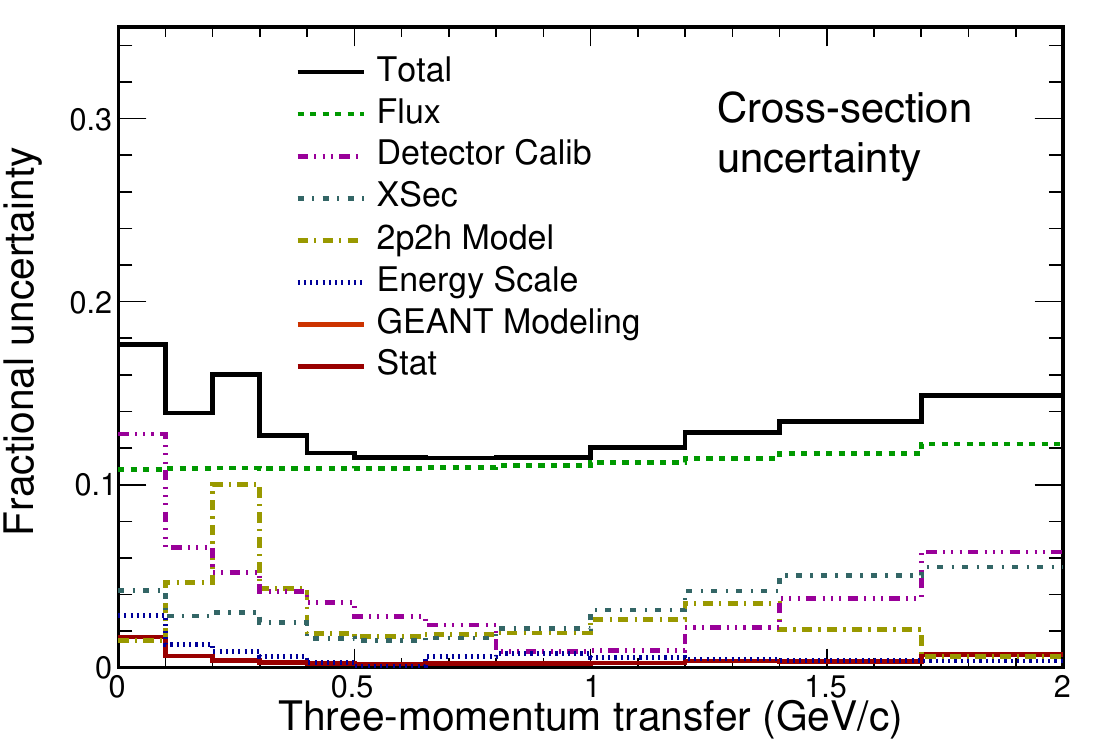}
\par\end{centering}
\caption{Fractional uncertainties on the cross section from 
systematic error sources vs. \qthree.
The histograms show the contributions of source categories to the total fractional uncertainty.}
\label{Fig06}
\end{figure}

\begin{figure}
\begin{centering}
\includegraphics[scale=0.47]{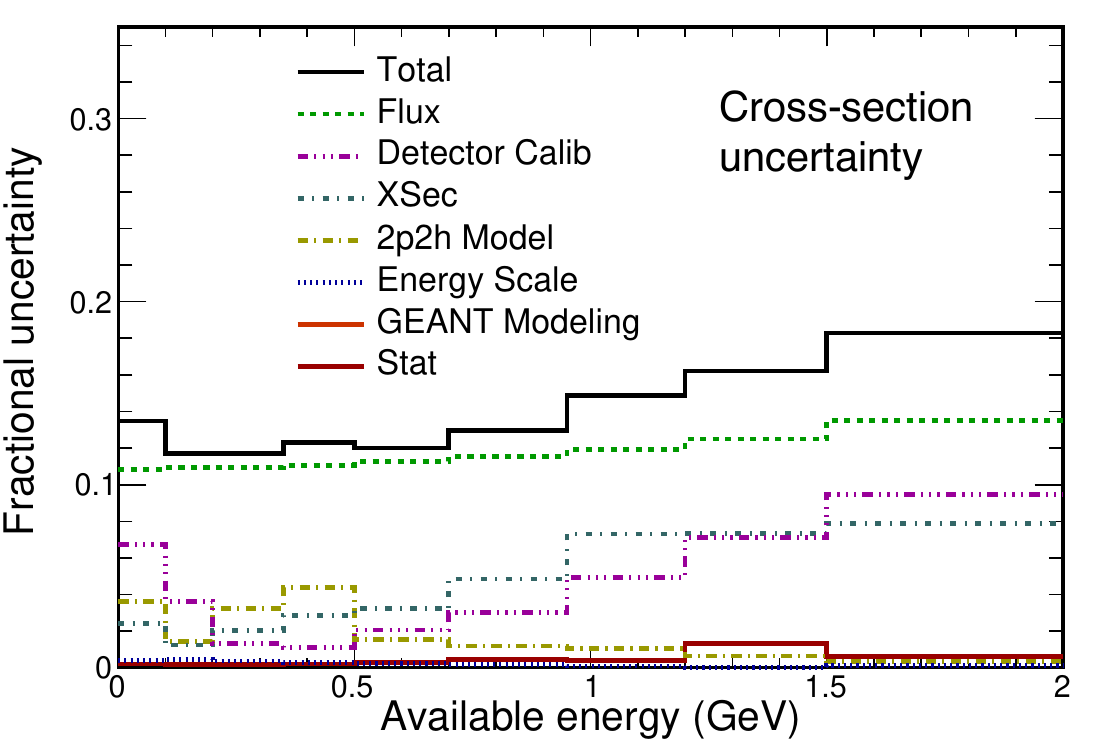}
\par\end{centering}
\caption{Fractional uncertainties on the cross section from systematic
error sources vs. \eavail.}
\label{Fig07} 
\end{figure}

The fractional uncertainties on the cross section arising from all sources of systematic error are
shown in Fig.~\ref{Fig06} in bins of \qthree\,, and in Fig.~\ref{Fig07}
in bins of \eavail.   In both figures, the flux (green dotted histogram) is the largest source of 
uncertainty in nearly all bins.   
The flux uncertainty is roughly uniform across
the phase space, staying within the range 10\,to\,14\%. 

Detector calibration (purple, dot-dash) gives the next largest uncertainty in low and high bins of
both \qthree\, and \eavail.  Uncertainties originating in modeling of neutrino-nucleon interactions (blue, dot-dash)
and the 2p2h process (olive-green, dot-dash) have significant presence in some portions of the phase space.
The total uncertainty (solid black) for projections onto bins in either variables is $< 19\%$ across the entire analysis domain.

Table~\ref{tab:Inclusive-Xsec-Weighted-Uncertainty} gives breakdowns of the
weighted average fractional uncertainties and correlations for the CC inclusive cross-section measurement.  
For the fractional uncertainties, the contribution from each source category  is averaged over all bins, with each bin weighted
by its cross-section content.  The bin-to-bin correlations from all sources of systematic uncertainty
are based on the difference between systematically-shifted simulations and the reference simulation, which is used
to calculate a total systematic uncertainty covariance matrix.   
The statistical covariance matrix is calculated separately
and the total uncertainty covariance matrix is taken as the linear sum of the systematic and statistical covariance
matrices.   More specifically, the weighted average fractional uncertainty, $\langle \delta \sigma/\sigma\rangle$,
is calculated as  $\left(\Sigma_{i}\, \sqrt{V_{ii}} \,\right)/(\Sigma_{i} \, \sigma_{i})$ 
where $i$ is a measurement bin, $V$ is the covariance matrix, 
and $\sigma_{i}$ is the measured double-differential cross section.   

\begin{table}
\begin{center}
\caption{\label{tab:Inclusive-Xsec-Weighted-Uncertainty} Fractional uncertainties and correlations for
the CC inclusive cross-section measurement, broken out by uncertainty source categories and averaged
over all bins.}
\begin{tabular}{ccc}
\hline 
\hline
Source of  & Weighted avg & Weighted avg \\
uncertainty &fractional uncertainty & correlation \tabularnewline
\hline 
Flux & 11\,\% & 1.0  \tabularnewline
2p2h model & 7.1\,\% & 0.6  \tabularnewline
Cross section model & 5.6\,\% & 0.2 \tabularnewline
Detector calibration & 3.7\,\% & 0.6  \tabularnewline
Energy scale & 0.9\,\%  & 0.6  \tabularnewline
Event statistics & 0.5\,\% & 0.4  \tabularnewline
\hline
Total & 17\,\% & 0.5  \tabularnewline
\hline 
\hline
\end{tabular}
\end{center}
\end{table}

The relative strength of correlations among the sources is indicated by the
weighted average correlation, $\langle \text{corr} \rangle$, 
whose value approaches 1.0 or 0.0
for strong or for neutral correlations, respectively.   
The values in Table~\ref{tab:Inclusive-Xsec-Weighted-Uncertainty} are calculated as 
$\langle \text{corr} \rangle = ( \Sigma_{i<j} \,C_{ij} \, \sigma_{i} \,  \sigma_{j} )/ ( \Sigma_{i<j} \, \sigma_{i} \,  \sigma_{j} )$ 
where $C$ is the correlation matrix, and where the indices $i$ and $j$ refer to 
different measurement bins~\cite{NOvA-CC-inclusive, Byron-Roe-Statistics}.

The flux is the leading source and it contributes
an average fractional uncertainty of 11\%.   The average correlation over all bins for the flux is 1.0,
indicating that this is mainly a normalization uncertainty.    The effect of the flux uncertainty can be alleviated in part by shape-only
comparisons of the measured cross section with predictions, and comparisons of this type are provided in subsequent sections.
Sizable correlations are also present for other uncertainty sources; however, these are subdominant relative to correlation with the flux.
The total systematic plus statistical uncertainty, 
calculated as a quadrature sum, is 17\%.   

\section{Double-differential cross section}
\label{sec:IncMeasure}

\begin{figure}
\begin{centering}
\includegraphics[scale=0.43]{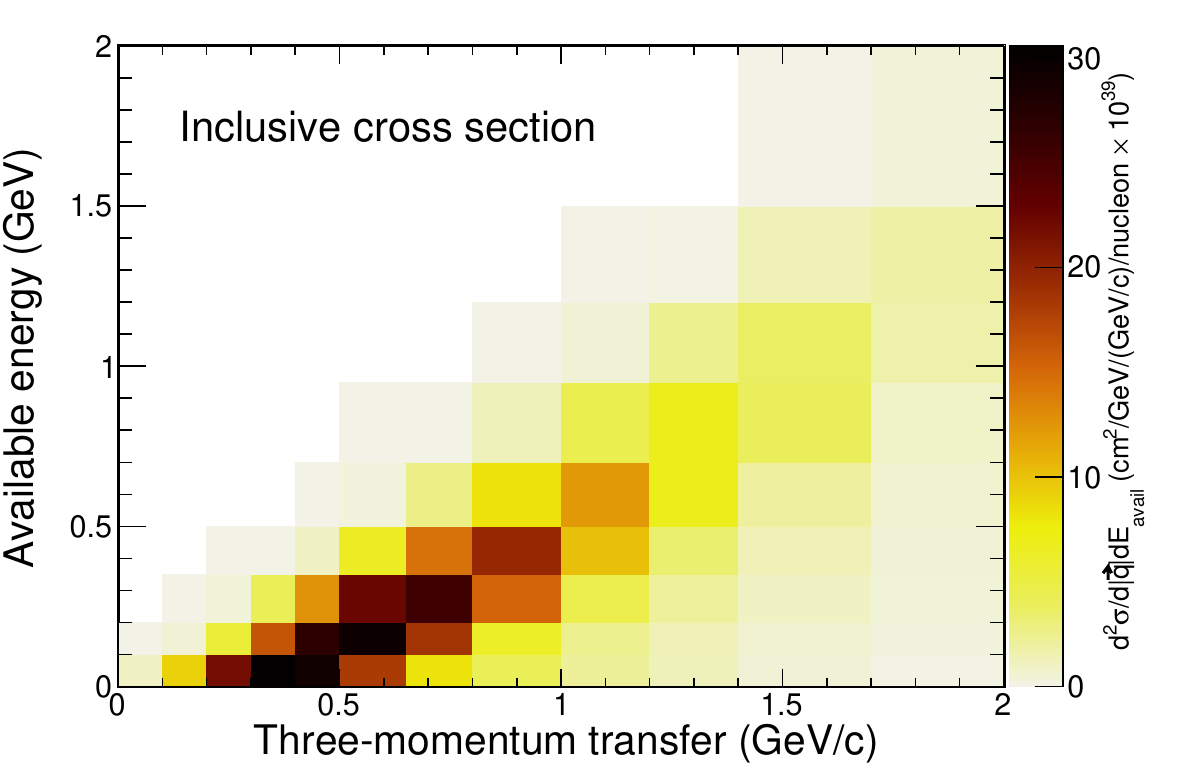}
\par\end{centering}
\caption{The flux-integrated, CC inclusive double-differential cross section
obtained by this analysis.}
\label{Fig08}
\end{figure}

\begin{figure}
\begin{centering}
\includegraphics[scale=0.43]{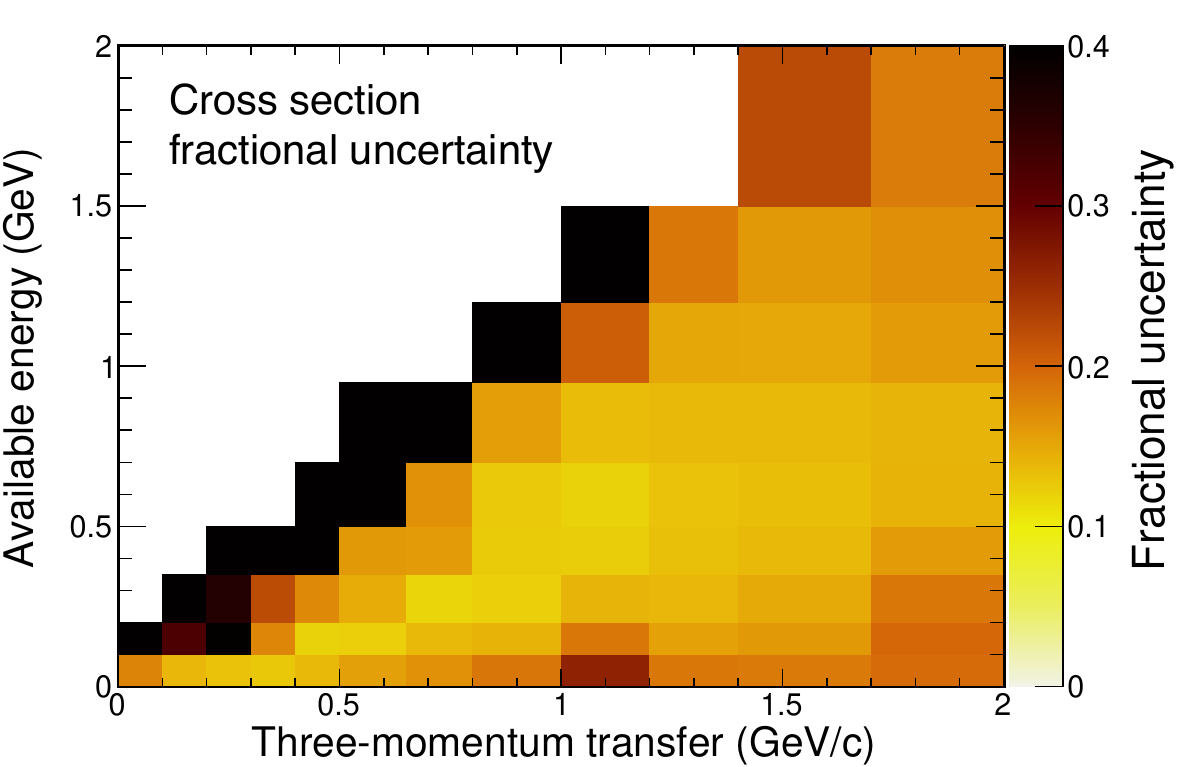}
\par\end{centering}
\caption{Bin-by-bin fractional uncertainties for the double-differential
cross section shown in Fig.~\ref{Fig08}.}
\label{Fig09}
\end{figure}

The distribution of signal events after unfolding and with 
correction for detection efficiency provides the foundation for the cross-section measurement.
Calculation of the flux-integrated, CC inclusive double-differential
cross section per nucleon was performed according to Eq.~\eqref{eq:XSec}.
The differential cross section thereby obtained 
is displayed in Fig.~\ref{Fig08} over the plane of \qthree\, and \eavail.
The cross sections reported by this analysis are based on the contents of the
68 bins with lowest fractional uncertainty (from the 72 bins displayed in Figs~\ref{Fig08} and \ref{Fig09}).

The cross section retains the ridge-like shape exhibited by the signal event distribution, with
the cross-section strength falling off as boundary regions of the analyzed phase space are approached.
The cross section peaks in the bin centered at 0.35 GeV/$c$ in \qthree\,
and 0.05 GeV in \eavail\, with a value of $3.35\times10^{-38}$ cm$^{2}$/(GeV/$c$)/GeV/nucleon.

The fractional uncertainty for cross-section bins of Fig.~\ref{Fig08}
is displayed in Fig.~\ref{Fig09}.
The uncertainty falls within 15\% to 20\% for most
 of the phase space, but becomes larger at the kinematic boundaries.
 Tabular summaries of the bin-by-bin cross-section and uncertainty values 
 are available in the Supplement~\cite{Supplement}.

\begin{figure}
\begin{centering}
\includegraphics[scale=0.42]{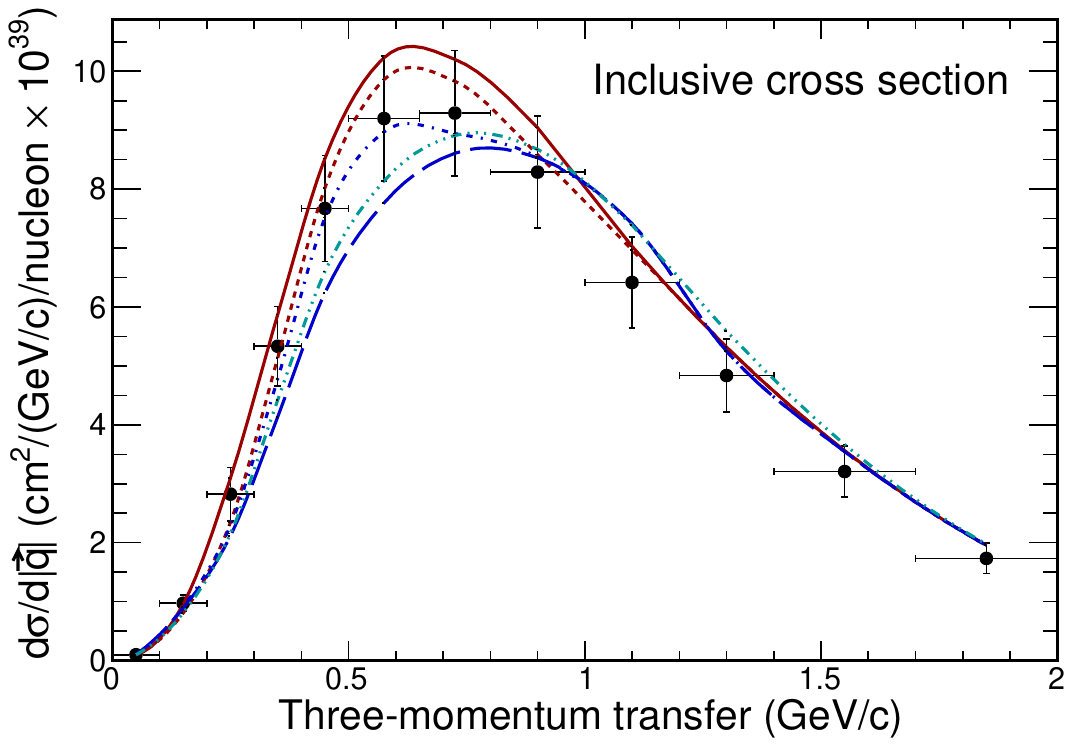}
\includegraphics[scale=0.42]{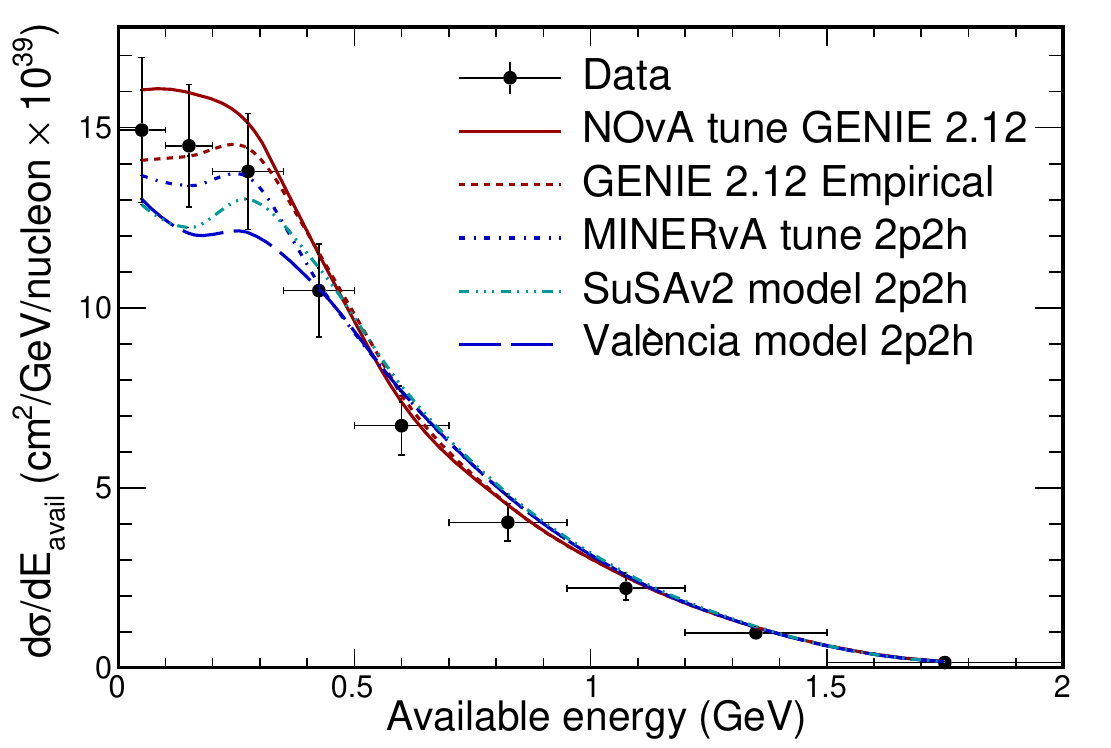}
\par\end{centering}
\caption{Inclusive single-differential cross sections $d\sigma/d$\qthree\, (top) and
$d\sigma/d$\eavail\, (bottom). The data (black crosses)
are compared to MC predictions using the five 2p2h models described in the text.}
\label{Fig10}
\end{figure}

Figure~\ref{Fig10} shows the single-variable
differential cross sections $d\sigma/d$\qthree\, (upper plot)
and $d\sigma/d$\eavail\, (lower plot).   Each of these is obtained
by integrating the double-differential cross section over the other variable.
The differential cross section for \qthree\, rises smoothly from
0 GeV/$c$, peaks at 0.65 GeV/$c$, and then decreases
roughly linearly with increasing \qthree\, beyond the peak.
The differential cross section for \eavail\, is largest from 0 GeV to 0.3 GeV 
and subsequently falls off rapidly.  The data (black crosses) are compared to simulations
rendered using the GENIE v2.12.2 neutrino event generator, in which the 2p2h models described in
Sec.~\ref{sec:Nu-Int-Modeling} have been employed.  The solid red curves show
predictions based on the NOvA cross-section tune used by the reference simulation of this analysis.   
Also shown are the predictions obtained with four other GENIE-based simulations, each of which uses a 
different 2p2h model.  The NOvA data tune gives a good representation of the data, while the GENIE Empirical and 
MINERvA data tunes under-predict the data through the peak regions.
Notably, the theory-based models of SuSAv2 and Val\`{e}ncia give even larger under-predictions,
both in the vicinity of the cross-section peaks and along the rising slope of $d\sigma/d$\qthree\, at low \qthree.

\begin{figure*}
\includegraphics [width = 14.5cm, height=9.32cm]{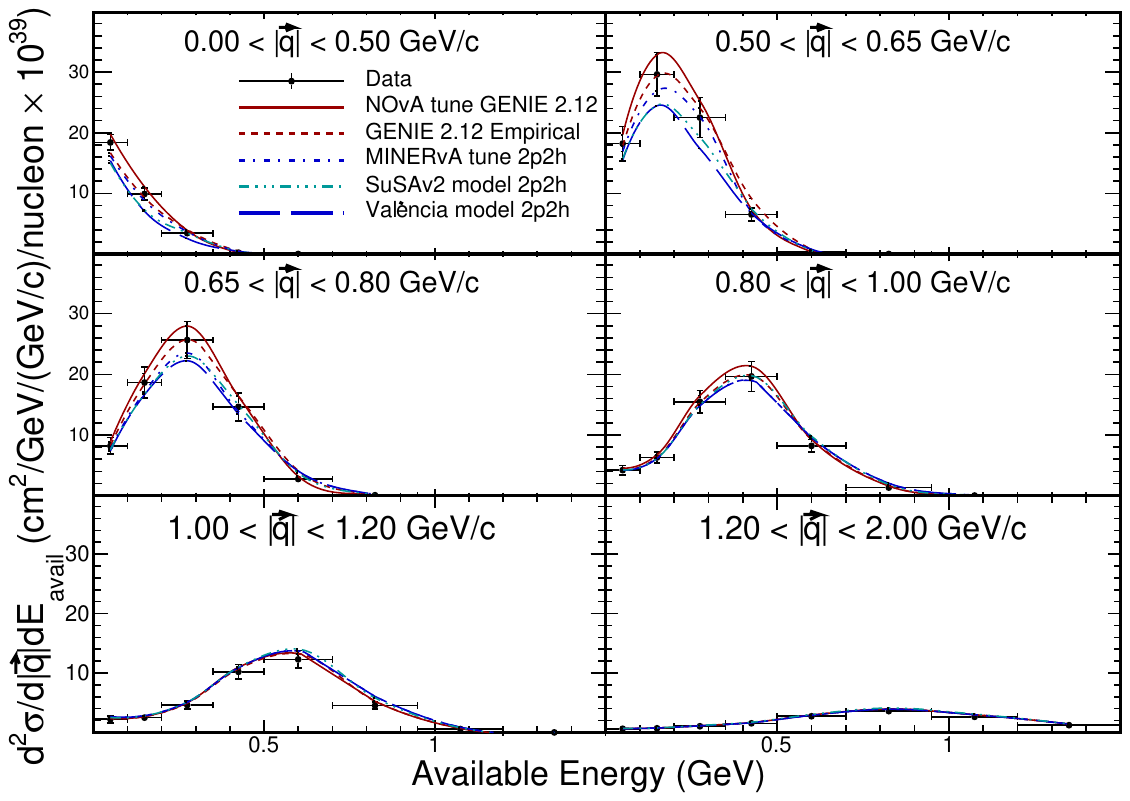}
\caption{The inclusive cross section in bins of \eavail, for six contiguous slices of \qthree. 
The data (crosses) are compared to predictions for \numu\,CC inclusive scattering that use five
different modeling implementations for 2p2h.}
\label{Fig11}
\end{figure*}

Figure~\ref{Fig11} shows the differential cross section 
in bins of \eavail\, for six contiguous ranges of \qthree\, wherein bins of Fig.~\ref{Fig08} have been merged.
The data (crosses) are compared with predictions obtained with the three neutrino-generator
tunes and two theory-based models considered by the analysis.   The predictions are in general agreement
concerning the evolutionary trend for 2p2h excitation through the six regions, however, differences in the 
absolute rate for 2p2h reactions are apparent.   As with the distributions of Fig.~\ref{Fig10}, 
the more-differential comparisons provided by Fig.~\ref{Fig11} indicate
shortfalls for predicted rates, especially those of the theory-based Val\`{e}ncia and SuSAv2 models.

\begin{figure}
\begin{centering}
\includegraphics[scale=0.46]{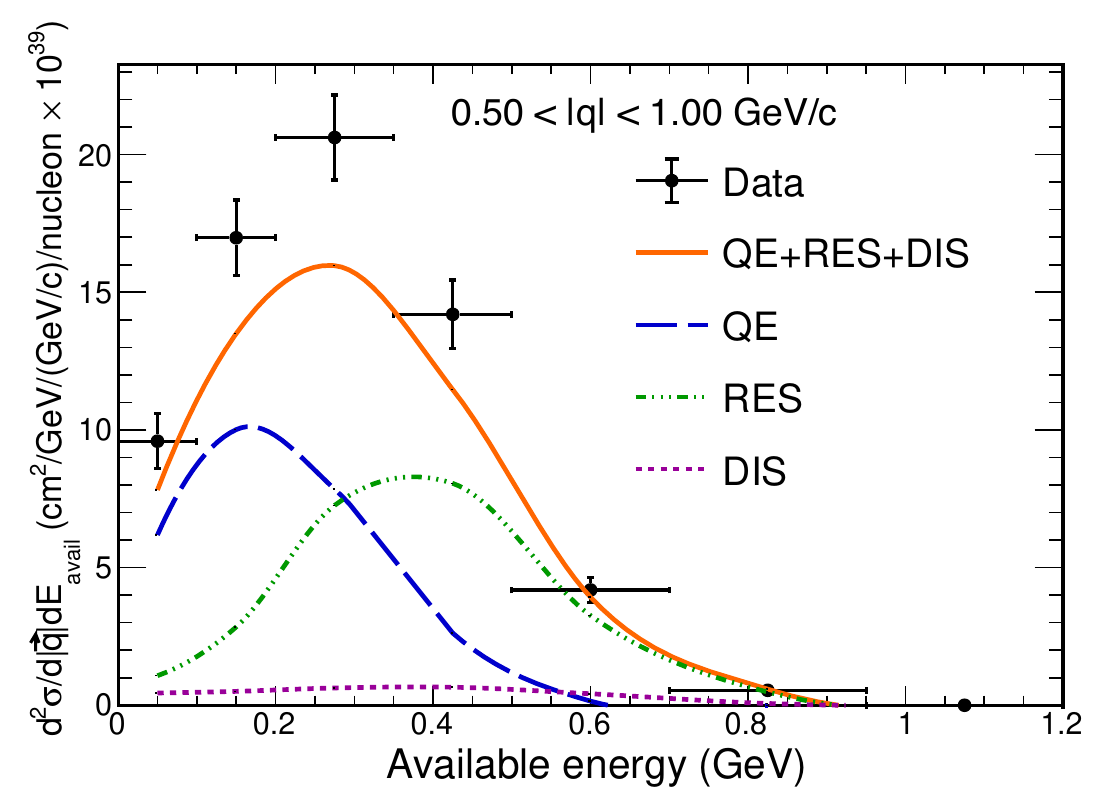}
\par\end{centering}
\caption{Cross section versus \eavail\, for the inclusive
cross section restricted to the range in three-momentum transfer where
the 2p2h contribution is estimated to be large. The data (crosses)
are compared to the sum of CCQE, RES, and DIS contributions as estimated
by NOvA-tuned GENIE v2.12.2 (solid orange curve).}
\label{Fig12}
\end{figure}

In Fig.~\ref{Fig11} the 2p2h contribution and the model spread are especially prominent 
in the range 0.5 $\leq$ \qthree\, $\leq$ 1.0 GeV/$c$.  
The measured cross section versus \eavail\, for this restricted range of \qthree\, 
is displayed in Fig.~\ref{Fig12}.  Here the data
are compared to the contributions from CCQE, RES, and DIS, but without 2p2h. 
The excess in the data is observed to be largest in the region of  \eavail\, that
lies between the CCQE and RES contributions, where the latter arises
predominantly from $\Delta(1232)$ resonance production. This
situation is as expected, for the appearance of 2p2h
in the kinematic region between elastic scattering and $\Delta$
production is well-established in electron-nucleus interactions~\cite{Katori-Martini-2018}. 
The trends in the data relative to the simulations as shown in Figs.~\ref{Fig11} and \ref{Fig12},
 are very similar to those reported by the MINERvA collaboration 
from CC interactions obtained using a wide-band
neutrino flux with spectral peak at 3.0 GeV~\cite{Rodrigues-2016}.

Table~\ref{tab:Inc2ModelChi2} shows 
the chi-square per degree of freedom ($\chi^{2}$/DoF) 
computed using the 2p2h predictions and data shown in Fig.~\ref{Fig08}, and using the full covariance matrix
described in Sec.~\ref{subsec:total-systematic}.
Columns 2 and 3 give the $\chi^{2}$ and $\chi^{2}$/DoF for comparisons involving both the cross-section shape 
and absolute rate using 68 DoF, while column 4 gives the 
$\chi^{2}$/DoF (67 DoF) for shape-only comparisons where the prediction is normalized to the measured cross section.

The lowest $\chi^{2}$ is obtained with the NOvA-tune prediction. 
This outcome is not surprising since
the tuning was done using neutrino data recorded by the NOvA Near Detector.
The next lowest $\chi^{2}$ is obtained with the GENIE Empirical model, with yet
larger $\chi^{2}$ values found for the other three models.  
Column 4 shows that the ranking by $\chi^{2}$ indicated by columns 2 and 3 remains the same
when the comparison is restricted to be shape-only wherein the predictions are normalized to the observed cross section. 
Interestingly, the covariances in the $\chi^{2}$ enable the SuSAv2 model to compare favorably with the MINERvA tune,
an outcome that cannot be readily inferred from comparing the predicted distributions for
the single-variable differential cross sections shown in Fig.~\ref{Fig10}.   Note that for 
the two theory-based 2p2h models, the $\chi^{2}$/DoF values worsen when the evaluations are restricted to shape-only.
For the Val\`{e}ncia model, restriction to shape-only doubles the $\chi^{2}$.

\begin{table}
\begin{center}
\caption
{Chi-squares with full covariances for predictions of GENIE-based simulations that use
different 2p2h models, compared to the measured CC inclusive double-differential cross section.   
Columns 2 and 3 give the $\chi^{2}$ and $\chi^{2}$/(68 DoF) for shape plus rate comparisons;
column 4 gives the $\chi^{2}$/(67 DoF) for shape-only comparisons.}
\medskip
\label{tab:Inc2ModelChi2}
\begin{tabular}{cccccc}
\hline
\hline
2p2h Model & $\chi^{2}$ &  $\chi^{2}$/DoF && \small{Shape Only}  \tabularnewline
\hline 
NOvA tune 2p2h & 270 &  3.96 && 3.25 \tabularnewline
GENIE Empirical & 550 &  8.08 && 7.36 \tabularnewline
MINERvA tune 2p2h & 746 &  11.0  && 11.7 \tabularnewline 
SuSAv2 2p2h  & 766 &  11.3  && 12.8 \tabularnewline
Val\`{e}ncia 2p2h  & 1501 &  22.1 && 46.0  \tabularnewline
\hline
\hline
\end{tabular}
\end{center}
\end{table}

The behavior of the 2p2h models with respect to each of the kinematic variables individually can be probed
by comparing predictions to the single-variable cross sections $d\sigma/d$\qthree\, and $d\sigma/d$\eavail\, 
displayed in Fig.~\ref{Fig10}.   Table~\ref{tab:IncModelChi2-q-and-Eavail} provides these
comparisons using $\chi^2$ with covariances.    The $\chi^2$/DoF values obtained with either of the 
cross-section distributions give similar rankings for the 2p2h models as is found with fitting to 
the double-differential cross section (Table~\ref{tab:Inc2ModelChi2}).   As previously, the $\chi^2$/DoF
are made larger for the theory-based models when the fitting is restricted to shape-only whereas the 
opposite trend is observed with the NOvA tune.

Figure~\ref{Fig12} indicates that 2p2h, together with CCQE, RES, and DIS, is a major component
of the CC double-differential cross section.
While the relative $\chi^{2}$ values of Table~\ref{tab:Inc2ModelChi2} clearly favor the NOvA tune for 2p2h,
they do not discriminate very strongly among the other 2p2h implementations.
These comparisons may be rendered less sensitive by the inclusion of regions of the analysis phase space 
where 2p2h has a small or negligible presence.    Identification of subregions of the 
\qthree\, versus \eavail\, phase space wherein the 2p2h contribution has a discernible
presence is therefore highly desirable.   Indeed, according to the models examined by this work,
the majority of 2p2h interactions occur in a single 
contiguous subregion of (\qthree,\,\eavail), the delineation of which is described in the next section.
This delineation enables examination of the \numu\,CC inclusive cross section of Fig.~\ref{Fig08}
to be focused on those bins that fall within the 2p2h-enriched subregion, with the remaining bins treated using
an overflow bin.   With the comparison of 2p2h predictions to data being made more localized in this way,
the testing of 2p2h models is different and perhaps more stringent than that 
provided by the full-phase-space comparisons of Table~\ref{tab:Inc2ModelChi2}.

\begin{table}
\begin{center}
\caption
{Comparisons of 2p2h-model predictions using $\chi^2$ with covariances, for the single-variable
cross sections of Fig.~\ref{Fig10}.  Columns 2 and 3 give $\chi^{2}$/DoF for fits to the total and shape-only
$d\sigma/d$\qthree\, (with 12, 11 DoF respectively);  columns 4 and 5 give $\chi^{2}$/DoF for fits
to $d\sigma/d$\eavail\, (with 9, 8 DoF).}
\label{tab:IncModelChi2-q-and-Eavail}
\begin{tabular}{c|c|c}
\hline 
\hline
2p2h &\qthree~Distribution & \eavail~Distribution \\
Model & $\chi^{2}$/DoF \,  \small{Shape Only} & $\chi^{2}$/DoF \,  \small{Shape Only}  \tabularnewline
\hline 
NOvA tune  & 2.45  ~~~~~  2.12  & 0.20 ~~~~~  0.12  \tabularnewline
GENIE Emp & 3.74  ~~~~~  3.65  & 0.67  ~~~~~ 0.68  \tabularnewline 
MINERvA  & 2.66 ~~~~~  2.84  & 4.32   ~~~~~ 4.84 \tabularnewline
SuSAv2   & 4.29  ~~~~~  4.90  & 5.72  ~~~~~  6.78  \tabularnewline
Val\`{e}ncia   & 5.34  ~~~~~  6.48 & 7.61 ~~~~~ 9.50   \tabularnewline
\hline 
\hline
\end{tabular}
\end{center}
\end{table}

\section{Estimation of 2p2h contribution}
A set of templates for event distributions over the plane of \qthree\, vs.
\eavail\, is assembled using the GENIE-based reference simulation of this analysis.
The templates are the predicted contributions from reaction 
categories that make up the total inclusive cross section.  The set
consists of the three major categories
CCQE, RES, and DIS, plus an additional low-population template ``Other" that
accounts for CC coherent scattering and purely leptonic inverse muon decay events.
If the reference simulation were completely accurate, then subtraction 
of the cross-section contributions represented by
the four templates from the measured double-differential cross section
would isolate an excess in the data that arises from 2p2h processes.

Data-based constraints on the event rate normalizations of the RES and DIS templates
are developed by defining a control sample which is nearly devoid of CCQE and 2p2h events.
Events in the data control sample satisfy at least one of the following two criteria:
{\it (i)}  The event has, in addition to the muon track, a particle prong of length $>$ 100 cm
(see Sec.~\ref{sec:Reco-and-Selection}).
{\it (ii)}  The event has three or more reconstructed prongs 
(in addition to the muon track) that emerge from the primary vertex.

The capability to modify the distribution shapes predicted by the templates is introduced by
dividing the analysis phase space and the templates 
into three regions of \qthree\,\,denoted I, II, and III,  wherein the RES and DIS template normalizations
are matched to the control sample.   The region boundaries are chosen
as ones that make optimal use of the control sample.  
{\it (i)} Region I is \qthree\,\,$\leq$\,\,1.2 GeV/$c$; this region constrains 
the RES normalization (see Fig.~\ref{Fig03}b).
{\it (ii)}  Region II is the intermediate region: 1.2\,$<$\,\,\qthree\,\,$<$\,1.4 GeV/$c$.
{\it (iii)}  Region III is the outer region: 1.4\,GeV/$c$ $\leq$\,\qthree;  it well-constrains the DIS normalization.

Simulation studies show that the NOvA detectors lack the resolution 
to distinguish between CCQE 1p1h scattering and the
manifestations that 2p2h interactions may have.    
One might envision, for example, that kinematic and/or ionization signatures from two reaction-induced protons 
(as in many 2p2h events) instead of just one (as in CCQE) could be the basis for subtraction of CCQE.  
Unfortunately this approach is not viable with the NOvA ND data.
Instead, the analysis bases its CCQE template and its normalization on 
the standard weak-interaction phenomenology used by GENIE v2.12.2 to model quasi-elastic scattering as related in 
Sec.~\ref{sec:Nu-Int-Modeling}.   Uncertainties are assigned to the parameters 
of this modeling as proposed by Refs.~\cite{GENIE-2015, RGran-archive},
with one exception: For the uncertainty associated with
the axial-vector mass, $M_{A}$, the reference simulation uses $M_{A}= 1.04 \pm 0.05$\,GeV~\cite{Meyer-2016}.

As remarked above, the relative contributions of RES and DIS are 
rather different in Regions I and III of the 
background control sample.
The approach adopted, after evaluating trial simulation runs,
is to adjust the RES and DIS normalizations in Region I via fitting 
to the control sample distribution in that region, 
while leaving the CCQE normalization at the nominal value assigned 
by the reference simulation.  
In the outer Region (III), the same procedure is used.
Then a final simultaneous fit to both Regions I and III is carried out 
wherein the normalization parameters starting values
are set according to the initial fit results.   The intermediate Region (ii) is the narrow region
of width 0.2 GeV/c that separates Regions I and III.   Since very little kinematic difference between
RES and DIS is predicted in this region, fitting was not done in Region II.   Instead, the RES and DIS
normalizations of RES and DIS are set to the average of the fit normalizations obtained in Regions I and III.
In this way, a degree of continuity is assured for RES and DIS template predictions over the entire
analysis phase space.  The contribution from the ``Other" template is quite small, 
and its normalization is also fixed at the nominal reference-simulation value.  
With the above-mentioned adjustments in place, each of the four templates (RES, DIS, CCQE, and Other)
are defined over the entire analysis phase space.

\begin{figure}
\begin{centering}
\includegraphics[scale=0.45]{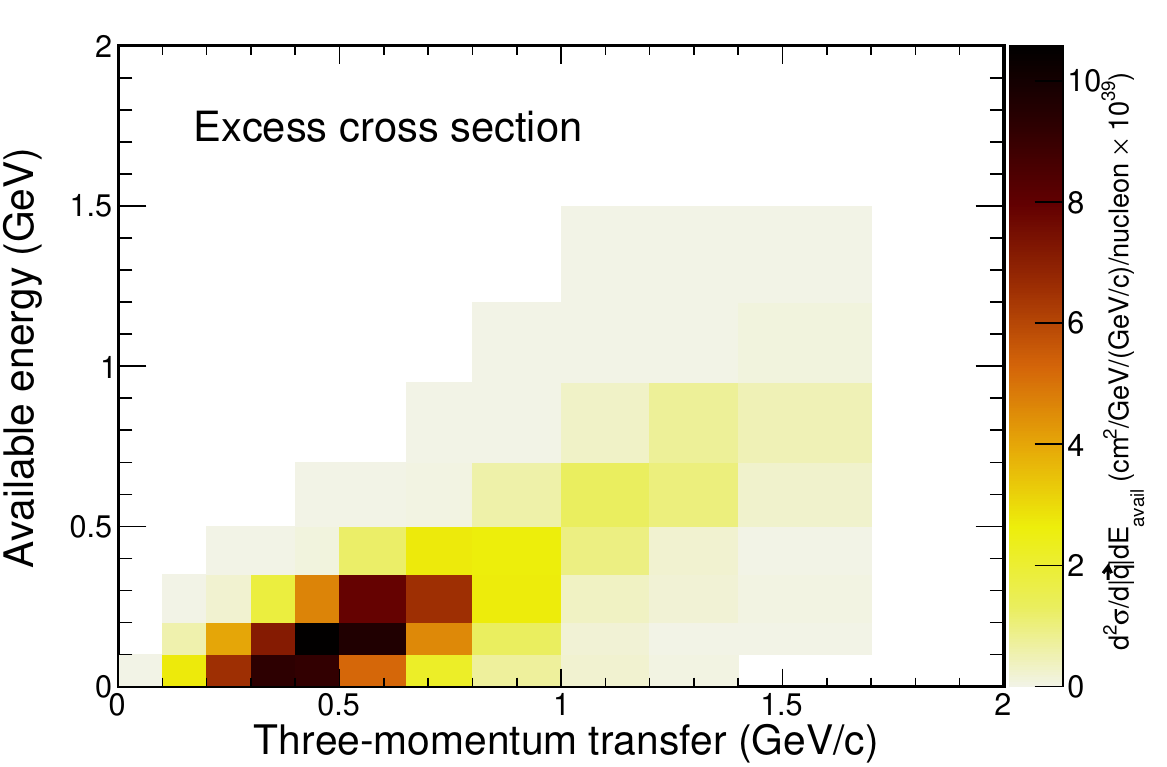}
\par\end{centering}
\caption{Double-differential cross section of excess events relative to GENIE-based estimation
of conventional CC neutrino-nucleon scattering. The cross section can be compared to the 
predictions of the 2p2h models and tunes.}
\label{Fig13}
\end{figure}

Figure~\ref{Fig13} shows the result of carrying out the subtraction of the sum over 
the four reaction templates from the distribution of selected signal events, and 
then converting the remaining event distribution into a cross section.
As described above, the populated bins show the data cross-section excess
relative to expectation derived from
the GENIE-based reference simulation, with RES and DIS contributions constrained by
the control sample, for CC neutrino-nucleon scattering
within nuclei modeled as a local relativistic Fermi gas.   
The subtraction of templates from the data gives rise to small
numbers of negative event counts appearing in four bins that are somewhat remote from regions 
with sizable event populations.  The affected bins are those with bin lower-edge values (in GeV/$c$ and GeV) 
as follows: (\qthree\,, \eavail\,) = (0.65, 0.5), (0.8, 0.5), (1.0, 0.7), and (1.4, 0.95).
The ratio of total negative to total positive event counts for the entire phase space is 0.029.  Among the four bins,
the last-mentioned contains more than 50\% of the negative event counts.   The contents of these four bins are 
set to zero prior to unfolding and the preparation of the cross section proceeds from that point.   
A separate unfolding study carried out for this ``excess" data sample gave the same outcome as described
in Sec.~\ref{sec:Backgrounds}.   On the basis of minimum $\overline{\text{MSE}}$, the sample was subjected
to two unfolding iterations.   The cross section finally reported and displayed in 
Fig.~\ref{Fig13} is obtained using these procedures.

Assuming that the contiguous, excess cross section of Fig.~\ref{Fig13} represents
2p2h interactions, then 2p2h is 23\% of the measured CC inclusive cross section.
The full reaction composition of the CC inclusive cross section according to the set of 
reaction templates finally obtained, is 23\% 2p2h, 38\% QE, 33\% RES, 5\% DIS, and 1\% Other.
This result differs from the composition predicted by the reference MC, which is 27\% 2p2h, 
38\% QE, 28\% RES, 6\% DIS, and 1\% Other.    The changes arise from the fit of 
RES and DIS templates to the data control sample, which calls for an 18\% increase in the RES
contribution (with consequent decrease in 2p2h) relative to the reference MC estimate.   These
changes are compatible with a total uncertainty of $\sim$ 20\% indicated for $\nu$-nucleus
RES production in NOvA data~\cite{Adjust-Models-2020} arising from uncertainties
 inherent to the Rein-Sehgal treatment of the $\Delta(1232)$ resonance and of higher-mass $N^*$ states, 
 including imprecise knowledge of the effective axial-vector mass and of low-$Q^2$ suppression.

The 2p2h cross section of Fig.~\ref{Fig13} is largest between 0.40 and 0.50 GeV/$c$ in \qthree\,
and between 0.10 and 0.20 GeV in available energy,  with a value
of $(1.1 \pm 0.3)\times10^{-38}\,\text{cm}^2\text{/(GeV/$c$)/GeV/nucleon}$. 
Smaller contributions are indicated at larger \qthree\, and
\eavail\, values.  The bin-by-bin cross-section 
fractional uncertainty is 28\% to 86\% in the region 0.3\,$\leq$\,\qthree\,$\leq$\,0.8\,
GeV/$c$ and 0.0\,$\leq$\,\eavail\,$\leq$\,0.35\,GeV.  

The per-bin cross-section fractional uncertainties for the extracted 2p2h signal arise from the same
source-of-error categories that characterize the CC-inclusive sample measurement, namely those
described in Sec.~\ref{Sec:Systematics} and summarized in Figs.~\ref{Fig06} and~\ref{Fig07}.
The relative contributions, however, are somewhat different.    For bins that contain the bulk of 2p2h rate, namely
0.3\,$\leq$\,\qthree\,$\leq$\,0.8\,GeV/$c$ and 0\,$\leq$\,\eavail\,$\leq$\,0.35\,GeV, uncertainties from the flux, 
CC cross-section modeling, and 2p2h modeling are comparable and fall in the range 10\,to\,40\%.   Uncertainty
from detector calibration generally falls below this range but becomes sizable (40\,to\,48\%) for \eavail\, exceeding 0.35\,GeV.
These rather large uncertainties are inherent to subtracting a large and partially unconstrained background
in order to estimate a signal which is roughly three times smaller.   Additionally, uncertainties allotted to parameters 
of the conventional phenomenology used for the CCQE background subtraction are necessarily conservative 
in order to cover modeling uncertainties associated with that reaction channel in a predominantly carbon medium.

\begin{figure*}
\includegraphics[width = 14.5cm, height=9.32cm]{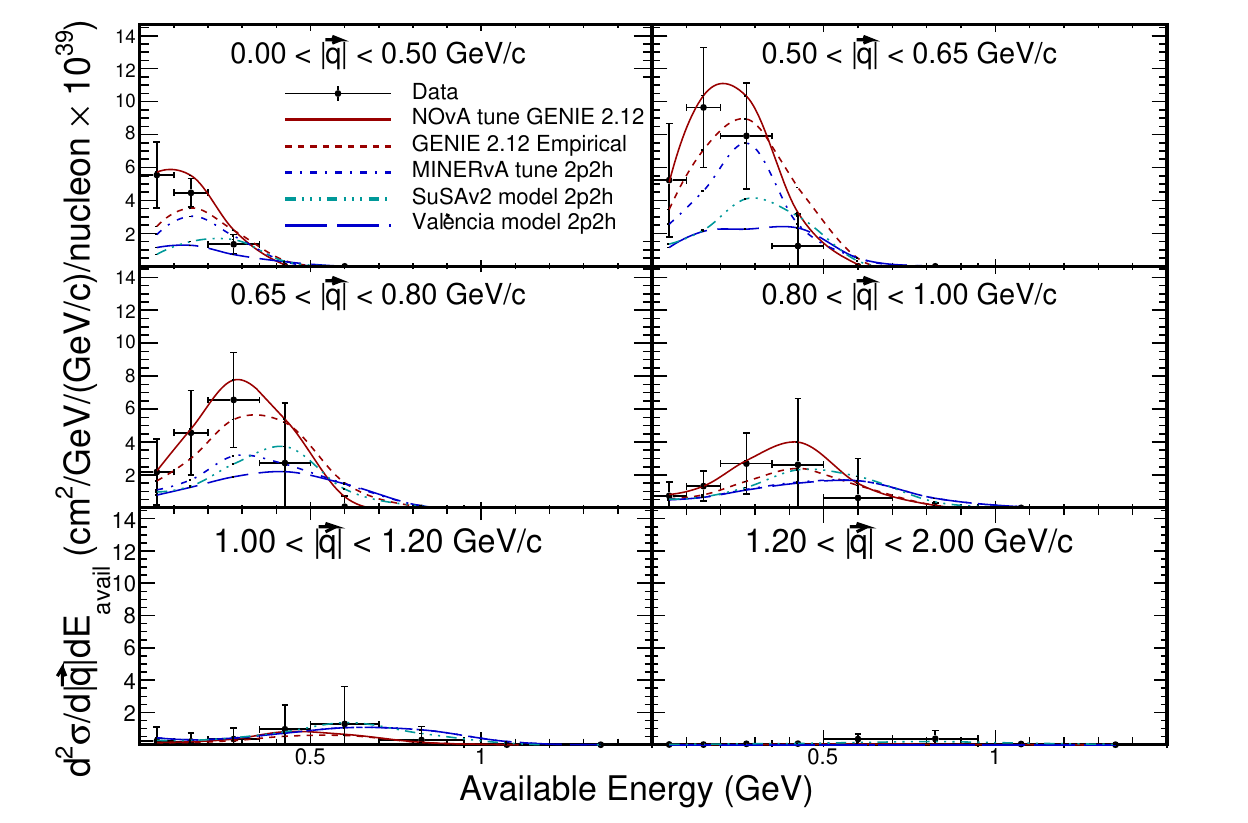}
\caption{The estimated 2p2h cross section of Fig.~\ref{Fig13} displayed
in bins of \eavail\, for six contiguous slices of \qthree.}
\label{Fig14}
\end{figure*}

Figure~\ref{Fig14} shows the data of Fig.~\ref{Fig13} plotted in bins of \eavail\, 
for six contiguous slices of \qthree.   The excitation pattern as a function of increasing \qthree\, follows
the trend predicted for the 2p2h contribution to the CC inclusive data, as displayed in Fig.~\ref{Fig11}.
The uncertainties on the data points are large throughout.   Nevertheless, discrepancies can be seen with 
cross-section rate and shape between the data (crosses) and the predictions of the SuSAv2 model (teal, dot-dot-dash curve)
and of the Val\`{e}ncia model (blue, long-dash curve).

\begin{figure}
\begin{centering}
\includegraphics[scale=0.42]{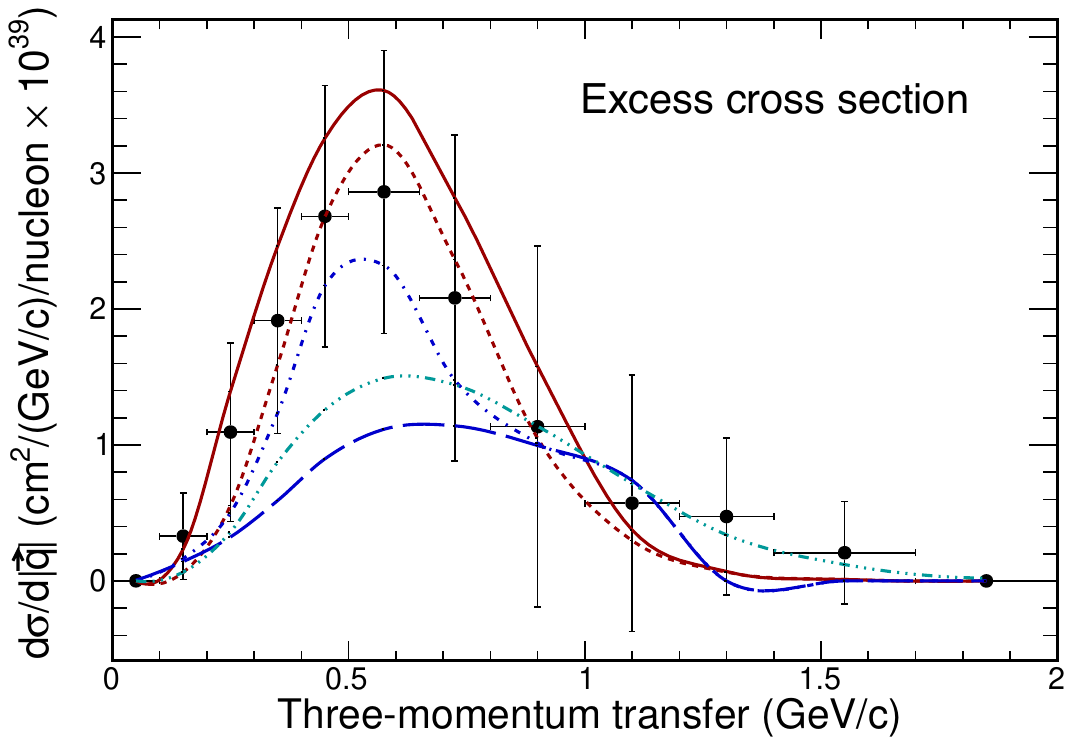}
\includegraphics[scale=0.42]{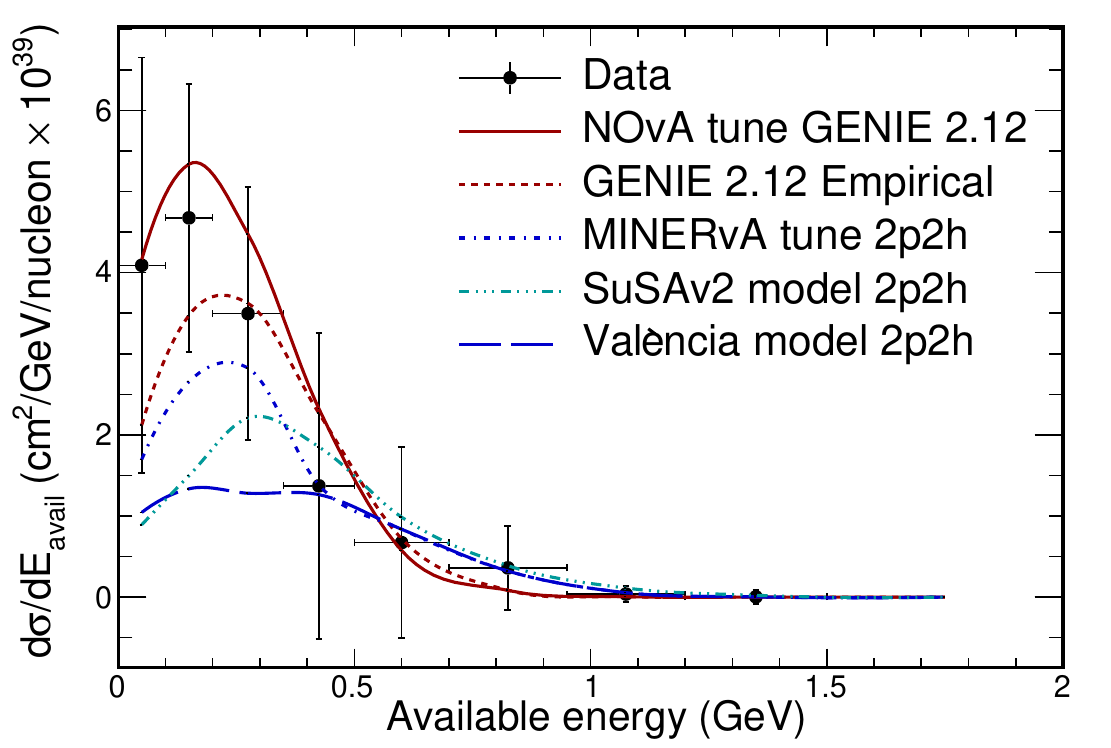}
\par\end{centering}
\caption{Distribution of the excess cross section 
projected onto \qthree\, and \eavail\,
and compared to 2p2h-model predictions.}
\label{Fig15}
\end{figure}

Figure~\ref{Fig15} displays the single-differential cross sections
in \qthree\, and in \eavail\, for the 2p2h contribution, and compares them to predictions
from the 2p2h tunes and models.   In these projections, the predictions of the NOvA tune 
exceed the data points in many bins, 
while predictions from the other models sometimes or often fall below the data points.
The two theory-based models
give relatively broader and flatter distributions than do the data tunes.   In particular, the Val\`{e}ncia model
predicts a two-component nature for 2p2h.  The components are predicted to have distributions that are kinematically
separated, with one being more CCQE-like and the other being more like a $\Delta(1232)$ excitation.   The peaks of these
distributions project onto adjacent but different points on the \eavail\, axis of the lower plot in Fig.~\ref{Fig15},
giving rise to a net distribution that is distinctly flatter than those predicted by the other models.
The data do not favor this aspect of the Val\`{e}ncia model.

Table~\ref{tab:Extracted-2p2h-Normal} provides comparisons of the five implementations of 2p2h 
to the extracted 2p2h double-differential cross section.   
The comparisons use the full $\chi^{2}$ including covariances, 
calculated over 18 bins for which the fractional uncertainty per bin is less than 100\%.   The bins cover the ranges
0.1-1.0\,GeV/c in \qthree\, and 0.0-0.35\,GeV in \eavail\,\cite{Supplement}.
Once again, the NOvA tune for 2p2h gives the lowest $\chi^{2}$ for the prediction to the overall cross section (columns 2 and 3), 
however respectable $\chi^{2}$/DoF values are found for most of the models.   Stronger distinctions are afforded by
the shape-only comparisons displayed in column 4, wherein model predictions are normalized to the total extracted cross section.
Restriction to shape-only fitting worsens the $\chi^{2}$/DoF for
all models except the NOvA tune.   This outcome is consistent with the disparities between predicted shapes
and the data that are discernible in Fig.~\ref{Fig15}.

\begin{table}
\begin{center}
\caption
{Comparisons based on $\chi^2$ with covariances, of 2p2h models to 
the extracted 2p2h cross section over the plane of (\qthree\,, \eavail).
Columns 2 and 3 give the $\chi^{2}$ and $\chi^{2}$/(18 DoF) for shape-plus-rate comparisons, while
column 4 gives the $\chi^{2}$/(17 DoF) for shape-only comparisons.}
\smallskip
\label{tab:Extracted-2p2h-Normal}
\begin{tabular}{cccccc}
\hline 
\hline
2p2h Model & $\chi^{2}$ &  $\chi^{2}$/DoF && \small{Shape Only}  \tabularnewline
\hline
NOvA tune 2p2h & 15.8 &  0.88 && 0.72 \tabularnewline
GENIE Empirical & 31.1 &  1.73 && 2.77 \tabularnewline
MINERvA tune 2p2h & 31.0 &  1.72  && 4.58 \tabularnewline
SuSAv2 2p2h  & 54.2 &  3.01  && 4.10 \tabularnewline
Val\`{e}ncia 2p2h  & 34.4 &  1.91 && 4.30  \tabularnewline
\hline 
\hline
\end{tabular}
\end{center}
\end{table}

\section{Predictions for \numu\,CC scattering in 2p2h-enriched region}
\label{sec:enriched-region}

Figure~\ref{Fig13} identifies regions of the analysis phase space 
populated by 2p2h interactions and enables a contiguous, 2p2h-enriched subregion to be defined.
For the purpose of model testing, the analysis defines such a region by restricting to
bins that contain $>$\,10\% of the cross-section value of the peak bin located at (\qthree,\,\eavail) = (0.4-0.5\,GeV/$c$, 0.1-0.2\,GeV).  
The selected bins form a contiguous region that extends from (\qthree,\,\eavail) = (0.1\,GeV/$c$, 0.0\,GeV) to (1.2\,GeV/$c$, 0.7\,GeV).
The binning is indicated by Fig.~\ref{Fig13} and Table~\ref{tab:Enriched-Bins}.
(Less restrictive choices, e.g., restriction to bins with $>$\,1\% or $>$\,5\% of the peak bin content, give similar results.)

As done previously for the comparisons of Table~\ref{tab:Inc2ModelChi2},
each of the five representations of 2p2h is used to predict the CC inclusive cross section,
and each prediction is compared to the measured double-differential cross section (Fig.~\ref{Fig08}) in
the 2p2h-enriched subregion.  The resulting $\chi^2$ values including covariances
are displayed in Table~\ref{tab:Inc2ModelChi2-2p2h-phase-space}.    
The $\chi^2$ values displayed in columns 2 and 4 show that the MINERvA and NOvA 
data tunes provide better matches to the data in the 2p2h-enriched region than do the other three 2p2h implementations.  Interestingly
the MINERvA tune compares well with the NOvA tune, even though it was adjusted to match data having a higher mean $E_{\nu}$ (3.0 GeV versus 1.86 GeV)~\cite{Rodriques-2016}.

\begin{table}
\caption{\label{tab:Enriched-Bins}Bin intervals of the 2p2h-enriched subregion.  Each column gives the \qthree\, interval
in GeV/$c$ (which is subdivided into bins)
associated with a given \eavail\, bin in GeV.}
\centering{}%
\smallskip
\begin{tabular}{cc|c|c|c|c}
\hline 
\hline
\eavail\, & 0.00-0.10 & 0.10-0.20 & 0.20-0.35 & 0.35-0.50 & 0.50-0.70\tabularnewline
\qthree\, & 0.10-0.80 & 0.20-1.00 & 0.30-1.00 & 0.50-1.00 & 1.00-1.20\tabularnewline
\hline 
\hline
\end{tabular}
\end{table}

\begin{table}
\begin{center}
\caption
{Comparisons using $\chi^2$ with covariances, of 2p2h models versus measured cross section over
the (\qthree\,, \eavail ) phase-space region enriched with 2p2h events.
Columns 2 and 3 summarize the shape-plus-rate comparisons for 22 DoF, while
column 4 gives the $\chi^{2}$/(21 DoF) for shape-only comparisons.}
\medskip
\label{tab:Inc2ModelChi2-2p2h-phase-space}
\begin{tabular}{cccccc}
\hline 
\hline
2p2h Model & $\chi^{2}$ &  $\chi^{2}$/DoF && \small{Shape Only} \tabularnewline
\hline 
NOvA tune 2p2h & 103 &  4.69 && 3.90 \tabularnewline
GENIE Empirical & 185 &  8.40 && 7.99 \tabularnewline 
MINERvA tune 2p2h & 84.4 &  3.83  && 4.11 \tabularnewline
SuSAv2 2p2h  & 177 &  8.04  && 9.15 \tabularnewline
Val\`{e}ncia 2p2h  & 347 & 15.8 && 18.6  \tabularnewline
\hline 
\hline
\end{tabular}
\end{center}
\end{table}

Since the $\chi^2$ for the NOvA 2p2h tune is the lowest for nearly all of the comparisons
provided by this work, one may wonder whether this outcome is the result of using the NOvA tune as the baseline model.   
To test this point, the analysis was re-run using, in turn,
each of the other 2p2h models as the baseline model.   In all four of these trial runs, the
ranking by $\chi^2$ is observed to be the same as reported in the rate-plus-shape and the shape-only comparisons 
of Tables~\ref{tab:Inc2ModelChi2},~\ref{tab:Extracted-2p2h-Normal}, 
and~\ref{tab:Inc2ModelChi2-2p2h-phase-space}.   Additionally, the $\chi^2$/DoF
values are similar (most are to within 0.5) to those obtained 
using the NOvA tune as baseline.    This outcome is to be expected, 
as dependence on baseline model is covered by the systematics treatment 
wherein the uncertainty spread is determined by running the entire analysis using different models.

In Table~\ref{tab:Inc2ModelChi2-2p2h-phase-space}, the two theory-based models give relatively larger chi-squares; 
the values in column four show this
trend to be more pronounced for the shape-only comparisons with the data.
The scattering amplitudes
invoked by the Val\`{e}ncia and by the SuSAv2 models are quite numerous 
and involve virtual pions, nucleons, higher-mass mesons, and baryon resonances.
The limited successes so far with this general approach suggest that important aspects of 2p2h 
still await an accurate theoretical characterization.  That said, the theoretical descriptions are likely to
improve in the near future, as further developments of the Val\`{e}ncia 
and the SuSAv2 models are in progress~\cite{Valencia-2020, SuSAv2-carbon-2022, Valencia-2024} 
and other approaches are being explored~\cite{Mosel-2024}.

\section{Conclusions}
\label{sec:Conclude}
This work reports a high-statistics measurement of the CC-inclusive double-differential
cross section $d^{2}\sigma/d|\vec{q}|dE_{\rm{avail}}$ for neutrino-nucleus interactions
of mean energy 1.86 GeV in a detector medium that is predominantly carbon but
includes heavier nuclei.   Differential cross sections over
the plane of~\qthree\,~and~\eavail\,~are presented in Figs.~\ref{Fig08}, \ref{Fig10}, and ~\ref{Fig11}.  
The selected event sample probes incident energies $1.0 \leq E_{\nu} \leq 5.0$\,GeV, 
a range of great importance to NOvA neutrino-oscillation measurements.   This $E_{\nu}$ range
lies above the sub-GeV to 1.5\,GeV region analyzed by T2K and MicroBooNE, while being
mostly below the 1 to 20 GeV region examined by MINERvA using its on-axis NuMI beam exposures. 
It covers the lower half of the high-flux plateau in the \numu~energy spectrum
planned for the DUNE experiment \cite{DUNE-2020}.   
This work extends the NOvA investigation of \numu\,CC inclusive scattering~\cite{NOvA-CC-inclusive}
in the above-stated $E_{\nu}$ range.
It also complements the experiment's recent measurement of the double-differential cross section
in muon kinematic variables of a 2p2h-enriched sample of CC low-hadronic-energy interactions~\cite{NOvA-low-Ehad}.

\smallskip

The CC-inclusive cross section receives contributions from 2p2h reactions wherein
more than one nucleon of a struck nucleus is involved in the interaction.  The
inclusive cross section for 2p2h is estimated from the data by subtracting 
template distributions for scattering on single nucleons predicted by a tuned version of GENIE v2.12.2
with normalization constraints for RES and DIS provided by a control sample.
The 2p2h cross section thereby inferred (Fig.~\ref{Fig13}) enables a restricted, contiguous region of phase space
enriched in 2p2h reactions to be identified.   Chi-square comparisons of GENIE-based predictions to \numu\,CC inclusive
scattering data  {\it (i)} using the full analyzed phase space (Table~\ref{tab:Inc2ModelChi2}), and 
{\it (ii)} restricting to the 2p2h-enriched region (Table~\ref{tab:Inc2ModelChi2-2p2h-phase-space}),
provide relative ratings for 2p2h models.
Chi-squares for predictions that use the SuSAv2 and Val\`{e}ncia 2p2h
models, and for predictions based on three different $\numu$-generator data tunes, indicate shortfalls with these
representations of 2p2h scattering.   The measurements of this work will facilitate the development 
of more accurate descriptions of \numu\,CC inclusive scattering
and of 2p2h reactions as is required by the long-baseline neutrino oscillation experiments.

\section*{Acknowledgments}
This document was prepared by the NOvA collaboration using 
the resources of the Fermi National Accelerator Laboratory (Fermilab), 
a U.S. Department of Energy, Office of Science, HEP User Facility. 
Fermilab is managed by Fermi Research Alliance, LLC (FRA), 
acting under Contract No. DE-AC02-07CH11359. 
This work was supported by the U.S. Department of Energy; 
the U.S. National Science Foundation; the Department of Science and Technology, India; 
the European Research Council; the MSMT CR, GA UK, Czech Republic; 
the RAS, the Ministry of Science and Higher Education, and RFBR, Russia; 
CNPq and FAPEG, Brazil; UKRI, STFC and the Royal Society, United Kingdom; 
and the state and University of Minnesota.  
We are grateful for the contributions of the staffs of the University of Minnesota 
at the Ash River Laboratory, and of Fermilab. For the purpose of open access, 
the author has applied a Creative Commons Attribution (CC BY) license 
to any Author Accepted Manuscript version arising.


\end{document}